\documentclass[12pt,preprint]{aastex}

\shorttitle{White dwarfs in 2MASS}
\shortauthors{Hoard et al.}

\slugcomment{Astronomical Journal, in press, 02/28/07}

\begin{document}

\title{Cool companions to white dwarf stars from the Two Micron All 
Sky Survey All Sky Data Release}

\author{D.~W. Hoard, S. Wachter}
\affil{Spitzer Science Center, California Institute of Technology, 
MS 220-6, 1200 E.\ California Blvd., Pasadena, CA 91125}
\email{hoard, wachter@ipac.caltech.edu}

\and 

\author{Laura K. Sturch\altaffilmark{1}, 
        Allison M. Widhalm\altaffilmark{2,3}, 
        Kevin P. Weiler\altaffilmark{4,5}, 
        Magaretha L. Pretorius\altaffilmark{6}, 
        Joseph W. Wellhouse\altaffilmark{1,3}, 
        Maxsim Gibiansky\altaffilmark{1}}
\affil{Spitzer Science Center, California Institute of Technology, 
MS 220-6, 1200 E.\ California Blvd., Pasadena, CA 91125}
\altaffiltext{1}{Department of Physics, Harvey Mudd College, 301 Platt Blvd., 
Claremont, CA 91711-5990}
\altaffiltext{2}{Department of Physics and Astronomy, University of 
Southern California, Los Angeles, CA 90089}
\altaffiltext{3}{Department of Astronomy, New Mexico State University, 
P.O. Box 30001, MSC 4500, Las Cruces NM 88003-8001}
\altaffiltext{4}{Department of Physics, Marquette University, 
P.O. Box 1881, Milwaukee, WI 53201}
\altaffiltext{5}{Department of Physics, DePaul University, Byrne Hall 211, 
2219 N.\ Kenmore, Chicago IL 60614}
\altaffiltext{6}{School of Physics \& Astronomy, University of Southampton, 
Southampton, SO17 1BJ, United Kingdom}

\begin{abstract}
We present the culmination of our near-infrared survey of the optically 
spectroscopically identified white dwarf stars from the McCook \& Sion 
catalog, conducted using photometric data from the Two Micron All Sky 
Survey final All Sky Data Release.  The color-selection technique, which 
identifies candidate binaries containing a white dwarf and a low mass 
stellar (or sub-stellar) companion via their distinctive locus in the 
near-infrared color-color diagram, is demonstrated to be simple to apply 
and to yield candidates with a high rate of subsequent confirmation.  
We recover 105 confirmed binaries, and identify 28 firm candidates 
(20 of which are new to this work) and 21 tentative candidates (17 of 
which are new to this work) from the 2MASS data.  
Only a small number of candidates from our survey have likely companion 
spectral types later than M5, none of which is an obvious L type 
(i.e., potential brown dwarf) companion.  
Only one previously known WD + brown dwarf binary is detected.
This result is discussed in the context of the 2MASS detection limits, 
as well as other recent observational surveys that suggest a very 
low rate of formation (or survival) for binary stars with extreme 
mass ratios.
\end{abstract}


\keywords{binaries: general --- infrared: stars --- stars: low-mass, 
brown dwarfs --- white dwarfs --- 
stars: individual(WD0014+097, WD0017+061, WD0018-267, WD0023+388, 
WD0027-549, WD0034-211, WD0041+092, WD0104-331, WD0116-231, WD0130-196, 
WD0131-163, WD0145-705, WD0145-221, WDJ0148-255, WD0205+133, WD0208-153, 
WD0232+035, WD0237+115, WD0248+601, WDJ0254-053, WD0255+009.2, WD0257+247, 
WD0258+184, WD0303-007, WD0302+621, WD0308+096, WD0309-275, WD0312+019, 
WD0347-137, WDJ0357+286, WD0354+463, WD0357-233, WD0357+081, WD0413-077, 
WD0416+272, WD0419-487, WD0429+176, WD0430+136, WD0458-662, WD0627+299, 
WD0628-020, WD0710+741, WD0718-316, WD0752-146, WD0800-533, WD0802+387, 
WD0805+654, WD0812+478, WD0824+288, WD0852+630, WD0858-220, WD0908+226, 
WD0915+201, WD0928+399, WD0933+025, WD0937-095, WD0949+451, WD0950+185, 
WD0956+045, WD1001+203, WD1004-178, WD1013-050, WD1015-173, WD1026+002, 
WD1027-039, WD1033+464, WD1036-204, WD1037+512, WD1042-690, WD1049+103, 
WD1055-072, WD1101+364, WD1104+044, WD1106+316, WD1106-211, WD1123+189, 
WD1126+185, WD1132-298, WD1133+358, WD1136+667, WD1141+504, WD1147+371, 
WD1156+129, WD1201+437, WD1210+464, WD1213+528, WD1214+032, WD1218+497, 
WD1224+309, WD1246+299, WD1247-176, WDJ1255+258, WD1254-133, WD1305+018, 
WD1307-141, WD1310-305, WD1314+293, WD1319-288, WD1330+793, WD1333+487, 
WDJ1340+604, WD1401+005, WD1412-049, WD1412-109, WD1415+132, WD1424+503, 
WD1433+538, WD1435+370, WD1436-216, WD1443+336, WD1458+171, WD1501+300, 
WD1502+349, WD1504+546, WD1517+502, WD1522+508, WD1541-381, WD1558+616, 
WD1603+125, WD1608+118, WD1610+383, WD1619+525, WD1619+414, WD1622+323, 
WD1631+781, WD1632-227.1, WD1634-573, WD1643+143, WD1646+062, WD1654+160, 
WDJ1711+667, WD1717-345, WD1729+371, WDJ1820+580, WD1833+644, WD1845+019, 
WD1844-654, WD1950+279, WD2009+622, WDJ2013+400, WDJ2024+200, WD2101-364, 
WD2108-431, WD2118-333, WD2131+066, WD2133+463, WD 2151-015, WD2154+408, 
WD2237+819, WD2256+249, WD2311-068, WD2317+268, WD2318-137, WD2326+049, 
WD2326-224, WD2336-187)}

\section{Introduction}

White dwarf (WD) stars play key roles in a wide variety of 
astrophysically important scenarios.  They represent not only the 
distant future of our own Sun, but are the endpoints in the evolution 
of the majority of stars in the Galaxy.  As the relic cores of normal 
stars, WDs reveal the outcome of stellar evolution processes, and 
expose material created during a stellar lifetime of nuclear burning 
to direct examination.  Yet, one of the most intriguing aspects of 
the observational study of WDs results simply from their role in 
facilitating the discovery of another type of object.

Detecting low mass stellar (or substellar)
companions to WDs offers many advantages compared 
to main sequence primaries.  
In the latter case, faint low mass 
companions are often hidden in the glare of the more luminous 
main sequence primary, and radial velocity variations are small 
and, therefore, difficult to detect.  
Since WDs are typically $\sim10^{3}$--$10^{4}$ times less luminous than 
main sequence stars, the brightness contrast compared to a 
potential faint companion is significantly reduced.  Also, 
the markedly different spectral energy distributions 
of the WDs and their low mass companions makes the detection and 
separation of the two components relatively straightforward even 
with simple broad-band multi-color photometry.  
Observational searches for cool companions to WDs that take 
advantage of these factors have been carried out during the past 
several decades (e.g., \citealt{probst83,zuckerman87,green00}).

In \citet[][henceforth, Paper I]{paperI}, we presented the 
Two Micron All Sky Survey \citep[2MASS;][]{skrutskie06} $JHK_{\rm s}$ 
photometry for all WDs from the \citet[][henceforth, MS99]{MS99} 
catalog that were contained in the 2MASS Second Incremental Data 
Release (2IDR).  We demonstrated that color-selection from the 
near-IR color-color diagram is an effective and efficient method 
for identifying candidate WD + low mass star binaries, via their 
near-IR excess compared to single WDs.  In Paper I, we recovered 
all 48 of the known, unresolved WD + low mass star binaries in MS99 
that were detected in the 2IDR sky coverage, and identified another 
47 new candidate binaries.  Our follow-up HST/ACS snapshot survey 
of candidates selected largely from Paper I has borne out the 
efficacy of this method, through the positive identification of 
a large fraction of resolved binaries with angular separations 
as small as $\sim0.05\arcsec$ (see the first results from this 
survey in \citealt{paperIII} -- henceforth, Paper III -- and 
discussion below).

In this work, we present the culmination of our correlation of 
the MS99 catalog with the 2MASS All Sky Data Release (ASDR).  
The results presented here are drawn from the final calibrated 
photometry for the objects from Paper I, as well as the remaining 
targets from MS99 that were not contained in the 2IDR sky coverage.  
Results for the subset of magnetic WDs were presented 
in \citet[][henceforth, Paper II]{paperII}, in which we did not 
find any strong binary candidates that could have been 
representative of the ``missing'' progenitor population of the 
magnetic cataclysmic variables (see discussion in Paper II and 
references therein).  In Paper I and, especially, Paper II, we 
utilized 2MASS photometry with relatively large uncertainties, 
which could have lead to the inclusion of several single WDs in 
our lists of binary candidates.  In this paper, we restrict our 
candidate sample to only those objects having 2MASS photometric 
uncertainties smaller than 0.1 mag in all three bands.

\section{The Data}

\subsection{Target Selection and Identification}
\label{s:phot_limits}

The source of our target sample is the MS99 catalog of optically 
spectroscopically identified WDs, which contains 2249 entries.
Recently, a large number of new WDs have been discovered 
(e.g., from the Sloan Digital Sky Survey; \citealt{eisenstein06}).  
However, we restricted our potential targets to just the sample 
in MS99 since, as described in Paper II, the new WDs discovered 
by the SDSS are almost all too faint to have been detected by 2MASS.
Our archival data were obtained from the 2MASS ASDR \citep{skrutskie06}, 
which includes $JHK_{\rm s}$ images and photometry 
covering 99.998\% of the sky.  
The photometric signal-to-noise ratio is $\geq10$ for the 
objects in the 2MASS Point Source Catalog (PSC) 
with $J\leq15.8$, $H\leq15.1$, and 
$K_{\rm s}\leq14.3$.  The overall detection limits of the survey are about
1 mag fainter in each band.  The PSC contains astrometry and photometry for 
almost 471 million objects.

Because many WDs have large (and often unknown) proper motions, 
their astrometry 
sometimes becomes unreliable only a few years after their discovery.  
Consequently, to identify each WD in 2MASS, 
we first searched the literature for 
each object.  In many cases we were able to locate published 
finding charts from the original, or a subsequent, 
identification of the WD;
for example, from the LHS atlas \citep{luyten79}, 
the Giclas proper motion survey and lists of suspected WDs 
(e.g., \citealt{G58} through \citealt{G80}), and the 
Montreal-Cambridge-Tololo survey \citep{lamontagne00}, to name 
only a few sources.  
These were compared with all available 
Digitized Sky Survey (POSS-I and POSS-II) 
and 2MASS images in order to identify the WD 
(after allowing for possible proper motion offsets, as described below).  

In some cases, finding charts were not available in the literature.
We then attempted to locate the WD using 
the most recent and accurate reported coordinates, proper motion, 
and/or blue optical color (e.g., by combining the POSS-II blue, red, 
and infrared images into a three-color composite)\footnote{We also 
checked all of our targets against the finding charts in the University 
of Arizona White Dwarf Database, at \url{http://procyon.lpl.arizona.edu/WD/}.  
However, this resource is still a work in progress and somewhat incomplete, 
and we found a number of instances in which the wrong star was identified 
as the WD (for example, often in the case of a wide common proper motion 
pair, the non-WD component was identified as the WD).  Thus, whenever 
possible, we preferred to use this resource only to confirm our independent 
WD identifications rather than using it as a primary source.}.
For example, when data from both surveys are available for a given WD, 
POSS-I and POSS-II provide images spanning decades, 
which makes proper motion based identification possible for high 
proper motion WDs.  When POSS-I images were not available, POSS-II 
and 2MASS images were compared for small changes in apparent position.  
Some Luyten Half-Second Survey objects, reported 
in \citet{bakos02}, fell into this category, and a combination of 
proper motion and precise coordinates was used for identification.  
Finally, we located the entry in the 2MASS PSC corresponding to the 
IR counterpart in the 2MASS images.  
Although this process is time-consuming, we estimate that simply 
using a coordinate match in the 2MASS PSC compared to MS99 would 
have resulted in up to 30--40\% of the WDs being misidentified or 
erroneously reported as undetected.

The MS99 catalog contains 2249 objects.  A number of the WDs have 
been subsequently reclassified as non-WDs (e.g., quasars) or 
non-existent (e.g., some WDs are listed twice in MS99 with 
different names, such as WD2009+397 = WDJ2013+400), leaving 
2202 viable targets.  Of these, we were unable to confidently 
recover the optical counterpart of the WD in 52 cases.  
For another 19 targets, even when we were able to recover the 
optical counterpart, the near-IR field proved to be too crowded 
to confidently identify the infrared counterpart.  This leaves 
2131 targets that we were able to confidently identify in optical 
and/or 2MASS images.  Of these, 656 were not detected by 2MASS 
(i.e., not listed in the 2MASS All Sky PSC), leaving us with a 
final sample of 1475 WDs detected in the 2MASS ASDR.

\subsection{Near-IR Photometry and Color-Color Diagram}

\begin{figure}
\epsscale{0.90}
\plotone{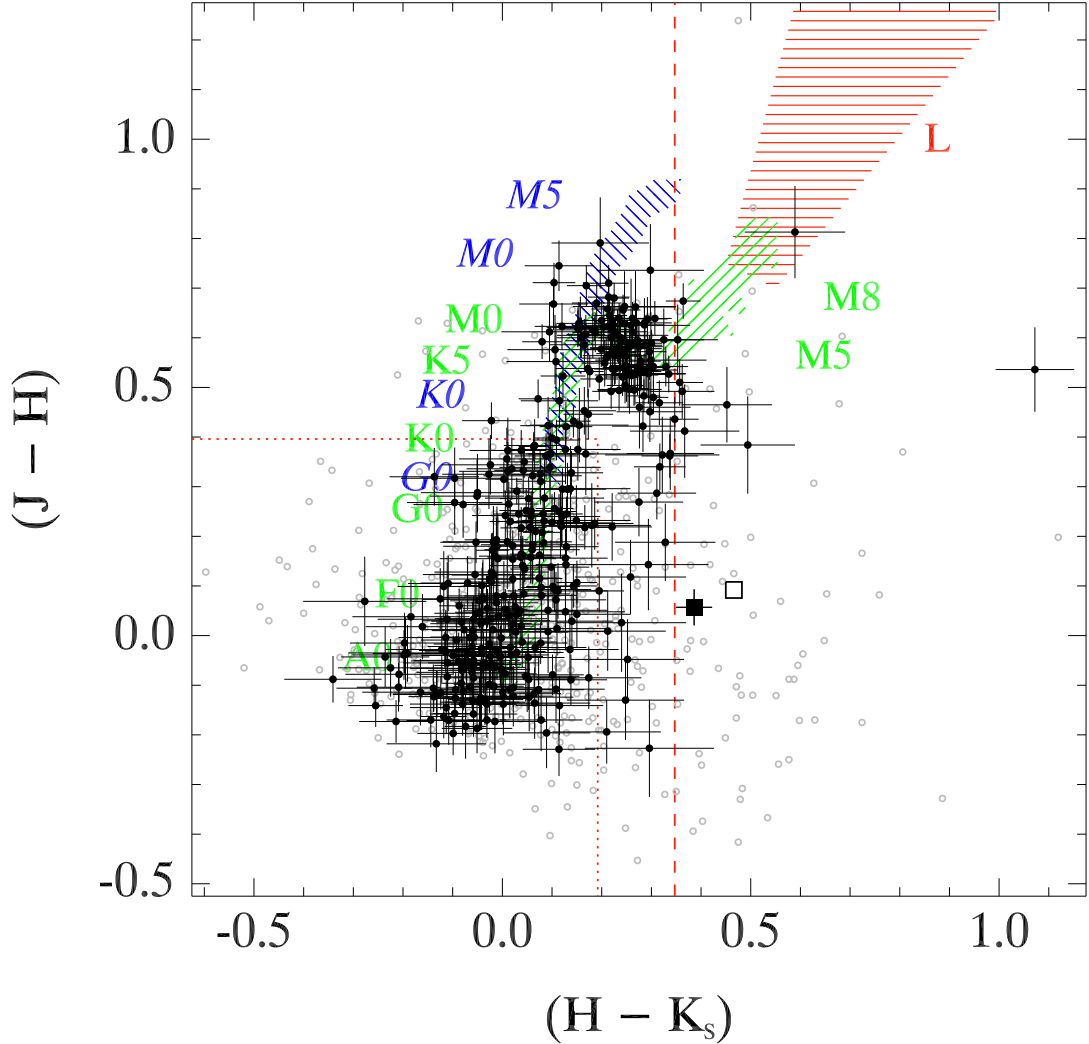}
\epsscale{1.0}
\caption{Near-IR color-color diagram for the WDs from MS99 that are 
detected in the 2MASS All Sky Data Release.  Filled black points 
with $1\sigma$ error bars are the Good detections.  Unfilled grey 
points are the Moderate detections.  (Poor detections are not shown, 
except as noted below.)  The squares are the two known WDs with 
circumstellar dust disks:\ WD2326+049 (G29-38; filled square, Good 
detection) and WD1729+371 (GD362; unfilled square, Poor detection). 
The cross-hatched regions show the loci of empirical mean 2MASS 
colors (\citealt{bessell88,gizis00,kirkpatrick00,hawley02}; 
transformed from other photometric systems using the relations in 
\citealt{carpenter01} when necessary) of the main sequence (///; 
green in the electronic edition),  giant branch 
($\setminus\setminus\setminus$; blue in the electronic edition) 
and L dwarfs (horizontal cross-hatches; red in the electronic edition).  
Spectral types are labeled at the correct $(J-H)$ value, but offset 
in $(H-K_{\rm s})$; spectral types of the main sequence and L dwarfs 
are labeled with roman font, while those of giants are labeled with 
italic font.  The dotted lines (red in the electronic edition) mark 
the boundary between ``normal'' WDs and red-excess WDs.  
The vertical dashed line (red in the electronic edition) marks the 
$(H-K_{\rm s})$ color of a dM5 star for comparison.
\label{f:ccd1}}
\end{figure}

As in Paper I, we graded the 2MASS photometry as ``Good'' 
(all $1\sigma$ photometric uncertainties $\leq0.1$ mag), ``Moderate'' 
(photometric uncertainty $>0.1$ mag in one or more of the three bands), 
and ``Poor'' (no formal photometric uncertainty in one or more bands, 
signifying a low signal-to-noise value more properly treated as an 
upper limit).  There are 417 Good detections, 475 Moderate detections, 
and 583 Poor detections.  Figure \ref{f:ccd1} shows the near-IR 
color-color diagram constructed from our data, including the 
$1\sigma$ error bars of the Good data.  The Moderate data are also 
shown (as unfilled grey points), but the Poor data are not plotted.  
Figure \ref{f:ccd2} is the same as Figure \ref{f:ccd1}, except that, 
for clarity, we have not plotted the error bars on the Good data.  
We restrict the remainder of our analysis and discussion to the 
Good data.

The color-color diagrams are similar in appearance to those presented 
in Papers I and II.  The primary difference compared to the color-color 
diagram in Paper I is the presence of a more populated and 
well-delineated ``bridge'' that extends almost vertically in the 
color-color diagram connecting the end of the WDs with colors similar 
to late main sequence stars to the bluer (and presumably single) WDs.  
This feature can be understood in the context of the color-color 
diagram for simulated WD + low mass star binaries shown in Figure 2 of 
Paper I, if one imagines how that figure would appear if the simulated 
binaries with $(H-K_{\rm s}) > +0.4$ were removed.   Thus, the bridge 
noted here is possibly identified as the tail end of the color 
distribution of candidate binaries containing WDs and the very lowest 
mass companions.  The tentative candidates from Paper I were drawn 
from this region of the color-color diagram, and (as will be discussed 
below) probably were contaminated by single WDs with red colors and/or 
larger photometric uncertainties.

\subsection{Binary Candidates}

The color selection boundaries used here are shown as dotted lines in 
the figures.  They correspond to $(J-H) \geq +0.396$ (equivalent to a 
dK0 star) and $(H-K_{\rm s}) \geq +0.192$.  All WDs outside (i.e., 
redward) of these boundaries were considered as candidate binaries.  
The reason for choosing the $(J-H)$ boundary is discussed in Paper I; 
however, the choice of the $(H-K_{\rm s})$ boundary is somewhat 
more arbitrary, 
and was guided primarily by our estimate of the middle of the 
``grey zone'' between the single WDs and the end of the simulated 
binary color distribution shown in Paper I (coincidentally, this 
value corresponds to the $H-K_{\rm s}$ color of a dM0 star).
Table \ref{t:phot} lists the 2MASS All Sky PSC photometry for all 
of the 154 WDs with near-IR excess selected in this fashion.

After compiling our list of WDs with near-IR excess, we re-examined 
the available literature for each of them in order to determine which 
ones are known binaries.  We then sorted each WD into one of three 
categories:\ confirmed, candidate, and tentative.  Confirmed binaries 
are those targets whose binary nature has been conclusively established 
in the literature, typically via spectroscopic detection of an 
unresolved low mass star, resolved imaging of the binary components, 
and/or detailed photometric modeling showing the presence of a low 
mass star spectral energy distribution after the WD is removed.  
Candidate binaries are those targets with near-IR excess reported, 
but not yet thoroughly investigated, in this work, Paper I, and/or 
other literature sources.  Tentative candidates are those targets 
for which there is substantial reason to doubt that the observed 
near-IR colors actually indicate the presence of a 
cool, low mass binary companion; 
for example, targets that fall into this category include DC (carbon) 
type WDs, possible WD+WD binaries, WDs with less certain optical 
identifications, etc.  In addition, any target selected as a 
candidate solely because it satisfies the $(H-K_{\rm s})$ criterion 
(but not the $J-H$ criterion) was categorized as tentative if 
its $(H-K_{\rm s})$ color was within $1\sigma$ of the $(H-K_{\rm s})$ 
color selection limit.  Not surprisingly, slightly more than half 
(12 out of 21) of the tentative candidates are drawn from the 
subset of WDs selected only because they satisfy the $(H-K_{\rm s})$ 
criterion.

Table \ref{t:binary} lists the binary status of each of the WDs 
from Table \ref{t:phot}, along with some supplemental information.  
The columns of Table \ref{t:binary} are as follows:\ (1) WD name, 
as listed in Table \ref{t:phot}; (2) binary status, as described 
above; (3) literature sources used to establish the binary status 
(``this work'' is not listed unless it is the only source or 
contains crucial evidence supporting the binary status -- see 
column 8); (4) and (5) estimates of the spectral types and angular 
separation of the binary components, if applicable (the 2MASS 
colors reported here are used to estimate a spectral type when 
possible; however, this estimate does not account for the presence 
of the WD, so the true spectral type is likely to be later than 
the nominal estimate); (6) the type of data used to establish the 
information in columns (4) and (5) (I = imaging, P = photomery, 
S = spectroscopy); (7) literature source(s) for the information 
in columns (4)--(6); (8) key for the presence of additional comments 
about individual WDs, listed in Appendix \ref{s:t2notes}.  
Multiple rows of data in columns (4)--(7) are sometimes available 
for a given WD.

\subsubsection{Anomalous Objects}
\label{s:anom}

Two objects stand out in the color-color diagram and deserve further 
explanation here (also see the corresponding notes in 
Appendix \ref{s:t2notes}).
WD1201+437 is the object located at extremely red $(H-K_{\rm s})$ 
color and is possibly a quasar.  
WD1517+502 is the only object located in the L spectral type region 
of the color-color diagram, but it is a known binary in which the 
WD's companion is a dwarf carbon star, not an L dwarf.

WD2326+049 formally qualifies for inclusion in our list of 
color-selected binary candidates, but has been excluded from the 
tables in this work because it is known to be surrounded by a 
dust disk (e.g., \citealt{graham90}, \citealt{tokunaga90}, 
\citealt{jura03}, \citealt{reach05}), and the presence of binary 
companions has been excluded down to angular separations of 
$a\geq0.9\arcsec$ and masses of $M_2\geq6M_{\rm J}$ \citep{debes05a}.  
For comparison with our known and candidate binaries, we have 
plotted the 2MASS colors of WD2326+049 and the other known WD 
with a dust disk (WD1729+371 = GD 362; \citealt{becklin05}) with 
special symbols in the figures.  The near-IR colors of these two 
WDs with dust disks are similar to each other.  
They satisfy only the $(H-K_{\rm s})$ color selection criterion, 
but fall outside the expected locus for WD + very low mass companion 
binaries shown in the simulation from Paper I.    
There are several WDs whose $1\sigma$ uncertainty ranges overlap 
with the WD + dust disk colors, but none whose nominal colors match.

\subsection{Number Statistics}
\label{s:stats}

Table \ref{t:stats} lists the total number of WDs with near-IR excess 
reported in this work, as well as the break-down into each of the 
three binary status categories used in Table \ref{t:binary} 
(i.e., confirmed, candidate, tentative).  We also show the numbers 
of WDs that are newly reported in this work and the current status of 
the binary candidates/WDs with near-IR excess reported in Paper I.  
Of the 154 objects reported in this work, 54\% are new, and the 
remainder were first reported in Paper I.  At the time of Paper I, 
only about one-half of our reported WDs with near-IR excess were 
already known to be binaries, but in the intervening years, more 
work has been accomplished in this field, and the fraction of 
confirmed binaries (as reported in the literature) from our total 
sample in this work is close to 70\%.  For example, 20 of the 
candidates from Paper I were subsequently confirmed by us to be 
binaries at small angular separations that are resolvable by HST+ACS 
(Paper III).  This leaves 28 firm binary candidates (20 of which 
are new to this work) and 21 tentative candidates (17 of which are 
new to this work).  

Of the 95 candidates from Paper I, 27 are ``missing'' from our 
current sample of WDs with near-IR excess.  Fifteen of these were 
removed because of better identification information\footnote{A few 
of these were noted independently in \citet{tremblay06}.} that leaves 
them as either undetected by 2MASS or with their identifications 
uncertain.  Another nine were excluded from the current candidate 
list because they have only Moderate or Poor 2MASS detections.  
The remaining three ``missing'' WDs are equally divided among being 
reclassified as something other than a WD, not satisfying the red 
selection criteria with the recalibrated All Sky photometry, and 
not being included in the All Sky PSC.  Of the 15 tentative 
candidates from Paper I, all but three have been excluded from the 
current candidate list because they have only Moderate 2MASS 
detections.

\section{Analysis and Discussion}

\subsection{Discussion of the Synthetic Photometry Method}

\citet{tremblay06} have recently suggested that a method involving 
calculation of synthetic photometry from WD model atmospheres is 
more efficient than our color-selection technique as a means to 
identify WDs with near-IR excess that are binary candidates.  
While the former method is undeniably effective and possibly has 
a lower ``false positive'' rate, it requires multi-band optical 
and near-IR photometry for each target, and is also more 
computationally and analytically complex than simple color 
selection from the pre-existing, homogenous 2MASS PSC.  
Consequently, we suggest that the latter method is, in fact, 
the more {\em efficient} method of identifying binary 
{\em candidates}, in terms of simplicity and ease of use 
(also see \S\ref{s:current_status}).  
The synthetic photometry method is likely more effective at 
providing subsequent confirmation and characterization of 
the near-IR excess in color-selected binary candidates 
(e.g., see discussion of WD0145-221 in \S\ref{s:missing}).

\subsection{Comparison with the Paper I Results}

\subsubsection{Current Status of Past Binary Candidates}
\label{s:current_status}

Overall, our ``success'' rate in finding binary candidates using 2MASS 
color selection that were subsequently confirmed as binaries is very high.
For example, in Paper I, we reported 47 new binary candidates (excluding 
the tentative candidates).  Eighteen of these have now been excluded 
from our sample for reasons other than simply having Moderate or Poor 
2MASS detections (see \S\ref{s:stats}).  Of the remaining 29, 28 have 
been confirmed as binaries by us in Paper III via high angular 
resolution imaging (20 objects) or using a method similar to that 
described by \citet{tremblay06} (8 objects).  
So, our confirmed success rate is 80\% (including the 48 already 
known binaries recovered as part of the original 95 candidates in 
Paper I), while our confirmed false positive rate (for targets that 
have been rejected as misidentifications in Paper I) is only 19\% 
(the remaining 1\%, one WD, is, as yet, neither confirmed as a binary 
nor rejected from the sample).  In this work, 68\% of the reported 
WDs with near-IR excess are recovered from already known or recently 
confirmed binaries.  We conclude that the near-IR color selection 
method is not only simple to apply, but also has a high success 
rate and correspondingly low rate of false positives (most of which 
are attributable to incorrect target identifications, not erroneous 
color-selection of single WDs).

Of the 15 tentative candidates reported in Paper I, only 3 remain 
in the sample reported in this work (with the rest having been 
excluded for having large uncertainties on their 2MASS photometry).  
One of these three has been confirmed as a binary, one has been 
``upgraded'' to a firm candidate due to a decrease in the 
uncertainties of its recalibrated photometry in the All Sky PSC, 
and the third remains a tentative candidate.  Eight of the 12 
excluded targets are now believed to be single WDs (e.g., as 
described in Paper II and \citealt{tremblay06}).  
There is a total of 21 tentative candidates reported in this work, 
12 of which were selected because they satisfy only the $(H-K_{\rm s})$ 
criterion (i.e., they are the targets identified with the simulated 
binaries containing the lowest mass companion stars).  
However, three of the targets selected from only the $(H-K_{\rm s})$ 
criterion are confirmed binaries (with the remainder of 
the $H-K_{\rm s}$-selected objects in the candidate class).
So, in this regard, we note that it is important when utilizing 
{\em any} method of selecting binary candidates to heed the warning 
implied by the label {\em tentative} candidate, but that rejecting 
these systems outright would result in the loss of a not insignificant 
fraction of true binaries\footnote{Incidentally, WD0518+333, which 
was identified as a tentative binary candidate in Paper I and is 
reported by \citet{tremblay06} as a single WD, is shown by our high 
angular resolution imaging to have a faint neighbor at a separation 
of $a\approx1.9\arcsec$ in 2005 (from a paper currently in preparation).  
If the neighbor is associated with the WD, then this is slightly 
below the expected 2MASS imaging resolution limit.  
If the neighbor is not associated with the WD (hence, does not 
share its proper motion), then the separation would have been 
$a\lesssim0.5\arcsec$ at the time of the 2MASS observation in 1998.  
Thus, this neighbor possibly contaminated the 2MASS photometry.  
We note that the known bright common proper motion companion to 
the WD (located $\approx8\arcsec$ northeast) is also resolvable 
into a close ($a\approx0.15\arcsec$) pair of stars that are 
approximately equally bright at $I$ band.}.

\subsubsection{Where Are the Binaries Containing Brown Dwarfs?}
\label{s:missing}

\begin{figure}
\epsscale{0.90}
\plotone{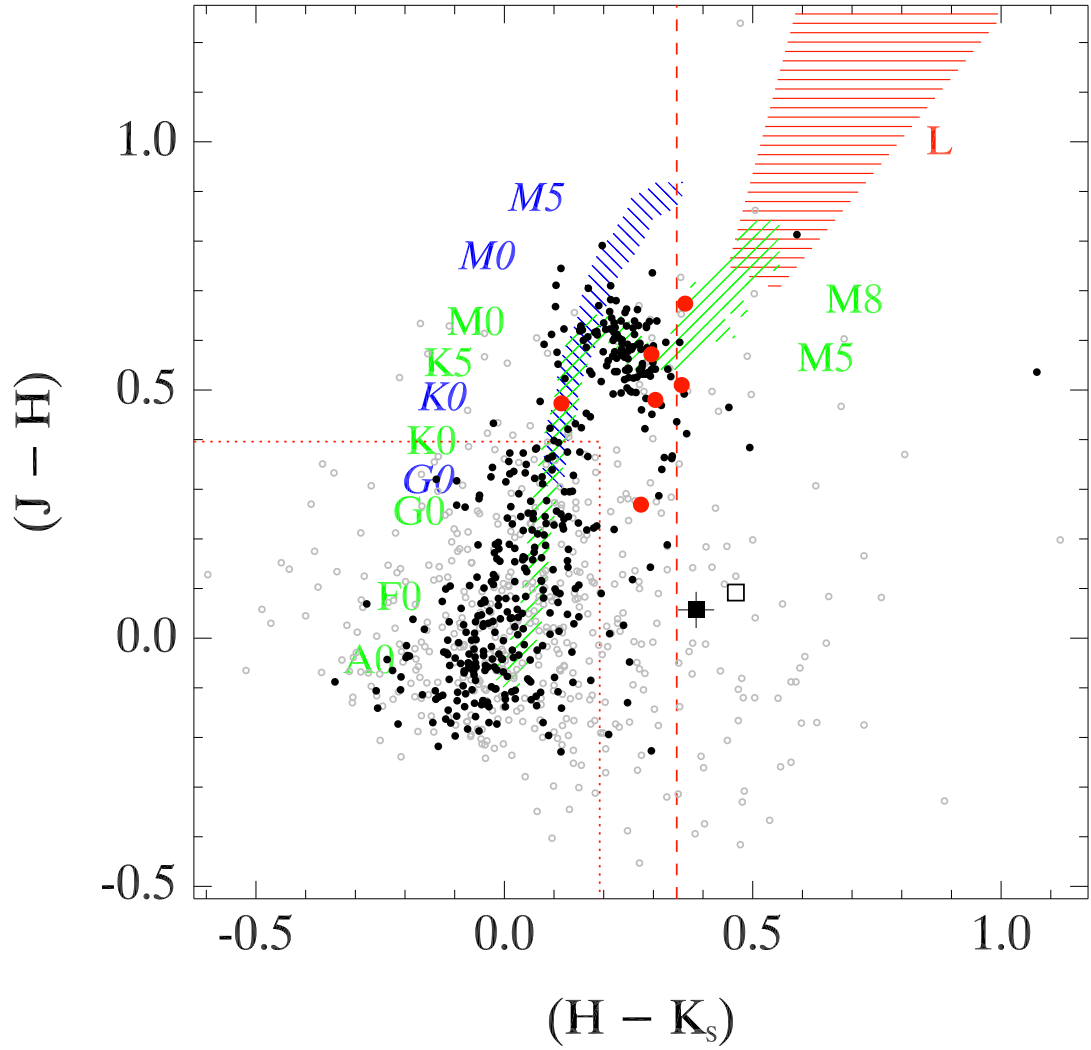}
\epsscale{1.0}
\caption{As in Figure \ref{f:ccd1}, but, for clarity, the error bars 
have not been plotted.  The six objects discussed in \S\ref{s:missing} 
are plotted with large symbols (red in the electronic edition).
\label{f:ccd2}}
\end{figure}

Since the near-IR color-color diagram shown in Figures \ref{f:ccd1} 
and \ref{f:ccd2} contains over 60\% more red-excess WDs than the 
color-color diagram from Paper I, a pertinent question is:\ why 
doesn't the current color-color diagram look more like the simulated 
diagram from Paper I?  Put another way, where are the binaries 
containing a WD and M5+ or L type (potential brown dwarf) companion?
Discounting WD1517+502 (see \S\ref{s:anom}), there are {\em no} observed 
candidates in the L spectral type region of the color-color diagram. 
Only eight targets in total (again excluding the two objects with 
anomolous colors -- see \S\ref{s:anom}) have nominal colors redward 
of the $(H-K_{\rm s})$ color of a dM5 star, although several times 
this many have colors within $1\sigma$ of this boundary (see vertical 
dashed line in Figures \ref{f:ccd1} and \ref{f:ccd2}).
Of the confirmed binaries in our sample, only six have estimated 
spectral types for the companion later than M5 (WD0145-221, WD0354+463, 
WD0419-487, WD0710+741, WD0752-146, and WD2151-015 -- see 
Table \ref{t:binary} and Figure \ref{f:ccd2}).  
Of these, only WD0145-221 has an L type companion (L6--7; i.e., 
a ``true'' brown dwarf); the rest have companion spectral types of 
M6--8.  However, the near-IR colors of WD0145-221 are most similar 
to an early dK star -- nowhere near the L spectral type region in 
the color-color diagram!  This object is at a distance of $d\approx39$ pc, 
so the companion star alone would not have been detected at the Good 
level by 2MASS (see below), and over 90\% of its $J$-band flux is 
attributed to the WD \citep{farihi04,farihi05b}.  
Although there is an observed excess in both $H$ and $K_{\rm s}$ 
bands compared to a single WD model, the observed slope of the spectral 
energy distribution from $H$ to $K_{\rm s}$ mimics 
that of the WD component (see Figure 1 in \citealt{farihi05b}).  
When combined with the almost complete lack of a $J$-band excess, the 
true nature of this binary is obfuscated in the color-color diagram.  
Although color-selected as a binary candidate, the brown dwarf 
nature of the companion is revealed only through additional analysis.  
It is possible then, that some binaries with brown dwarf companions 
are similarly ``hiding'' in the near-IR color-color diagram among 
the total sample of red-excess WDs.

However, we should also explore this issue from the point-of-view 
of the assumptions that went into the simulation from Paper I, 
compared to the real characteristics of the 2MASS survey and the 
expectations for the presence of WD + brown dwarf binaries.
In this context, part of the answer lies in the fact that the 
simulation in Paper I assumed that all of the simulated binaries 
are located at $d=10$ pc and that all of them are detectable by 2MASS.  
In fact, most ($\sim70$\%) of the WDs from Table \ref{t:phot} have 
distance estimates in the literature (e.g., compiled in Paper III; 
\citealt{holberg02,silvestri02,farihi05a,liebert05}), and we find 
an average distance of $d=150$ pc for our binary candidates, with 
distances ranging from $\approx5$--$700$ pc (plus one $7\sigma$ 
outlier at $d=1660$ pc).
Table \ref{t:dists} lists the absolute magnitudes of late main 
sequence spectral types, along with the corresponding maximum 
distance moduli for detection within the $S/N\geq10$ photometric 
limits of 2MASS (see \S\ref{s:phot_limits}).  
So, for example, stars later than M5--6 are only detected (at the 
Good level) if they are closer than $d\approx150$ pc (distance 
modulus of 5.88), while stars later than L6 are only detected if 
they are closer than $d\approx25$ pc (distance modulus of 1.99).  

The table also lists the mean photometry for the sample of single 
WDs used in the simulation from Paper I.  If we take these values 
as representative, then for a given WD + low mass star binary, 
the components will be approximately equally bright in the $J$ and 
$H$ bands for companion spectral types of L3--5 and L5--6, 
respectively.  The WD never exceeds the $K_{\rm s}$-band brightness 
of the companion at even the latest L spectral type.  
In the case of binaries containing equally bright components 
(in a particular photometric band), the combined absolute magnitude 
is 0.75 mag brighter than either component alone, allowing for 
detection out to a factor of $\sqrt{2}$ larger distance.  
However, as the companion spectral type becomes earlier or later, 
the combined photometry of the binary is rapidly dominated by the 
companion or the WD, respectively, and the maximum distance modulus 
for a Good detection in 2MASS rapidly converges back to the single 
star values listed in Table \ref{t:dists}.  
Thus, we should expect that binaries containing late-M or mid-L 
companions should be detectable at the Good level in 2MASS out 
to distances of $d\approx150$ pc or $d\approx(25)(\sqrt{2})=35$ pc, 
respectively.  While this does imply that there will be fewer 
detected binaries containing mid-L companions than late-M companions, 
it does not imply the almost complete non-detection of the former 
that we observe.  

To further test this theory, we
modified the simulation from Paper I to randomly assign a distance 
from a distribution equivalent to that for our current targets, and 
then reject simulated binaries that would not be detected in 2MASS 
at the Good level.  We also tried a number of different distance 
distributions, such as increasing the relative number of targets 
at a given distance, $d$, in proportion to $d^{2}$ between several 
minimum and maximum distances, or assuming a strongly peaked 
population at a specific distance that falls off for larger and 
smaller distances.  As suspected, although in all cases introducing 
some kind of distance dependence on the dectability of a simulated 
binary results in fewer late spectral type binaries in the simulated 
color-color diagram, there are still a substantial number of 
``detected'' binaries with late-M to mid-L companions, which are 
not present in large numbers in the observed color-color diagram.

We can also explore the effect of the observed relative numbers 
of stars as a function of spectral type (or mass).  
By default, the simulation from Paper I (which was constructed 
purely to illustrate the possible loci of near-IR color-color 
space occupied by WD binaries) considers a binary with an M type 
companion to be as likely as a binary with an L type companion.  
In reality, however, only about 35\% of the field stars within 
20 pc have spectral types of M5 or later and only about 5\% have 
spectral types of L0 or later; most ($\approx50\%$) have spectral 
types of M3--4 (e.g., as compiled in \citealt{farihi05a}; see 
their Figure 7).  Taken at face value (however, see below), this 
suggests an explanation for why we {\em do} observe numerous 
candidate binaries with near-IR colors equivalent to main sequence 
stars up to spectral type of M4--5.  Considering the distribution 
of field star spectral types, from our sample of 154 confirmed and 
candidate WD binaries, we might expect 50--55 to contain dM5+ type 
companions, $\approx8$ of which are L type companions.  
Approximately 50\% of the red-excess WDs with known distances are 
closer than $d\approx150$ pc (i.e., detectable if they contain a 
late-M companion); approximately 20\% are closer than $d\approx35$ pc 
(i.e., detectable if they contain a mid-L companion).
So, to first order, we might still expect to detect 25--30 binaries 
with dM5+ companions, of which 1 or 2 have mid-L companions.  
The latter is roughly in agreement with our observed sample, 
if we count WD0145-221 as the one detected WD + brown dwarf binary; 
however, the former is consistent only if we count essentially 
all of the binary candidates within $1\sigma$ of the $(H-K_{\rm s})$ 
color of a dM5 star as binaries containing dM5+ companions.

Of course, the situation is more complex than this, since numerous 
recent surveys for WDs in binaries with low mass companions have 
noted that there appears to be a mechanism 
that makes the existence of binaries containing a WD and very low 
mass (late-M through L type) companion less likely than that of a 
WD + mid-M or earlier companion, in excess of the relative numbers 
of low mass field stars of these spectral types.
The well-known ``brown dwarf desert'' describes an observed dearth 
of solar-type stars (i.e., WD progenitors) in binaries with brown 
dwarf companions at separations $r\lesssim5$ AU, compared to the 
frequency of low mass star + brown dwarf and brown dwarf + brown 
dwarf binaries (e.g., \citealt{marcy00,grether06}).  
The brown dwarf desert might extend out to at least many hundreds of AU, 
although there is some evidence that the desert does not extend past 
separations of $r\gtrsim1000$ AU (\citealt{gizis01}; however, also see 
the apparently contradictory results from 
\citealt{farihi05a} -- discussed below -- and \citealt{mccarthy04}).  
An additional factor that could contribute to a lack of WDs with 
(close) brown dwarf companions is the possible destruction or 
outspiraling of pre-existing low mass companions during the post-main 
sequence evolution of the WD progenitor (described in Paper III and 
references therein).  

At distances of 5--400 pc, and assuming an angular resolution limit 
of $\sim2\arcsec$, the binaries in our sample have likely separations 
of $r\lesssim10$--$800$ AU.  \citet{farihi05a} performed a search 
of 261 WDs sensitive to low mass companions at separations 
of $r\sim100$--$5000$ AU and an additional search of 86 WDs sensitive 
to low mass companions at separations of $r\sim50$--$1100$ AU.  
They detected {\em no} brown dwarf companions, implying that the 
fraction of WD + brown dwarf binaries, even at large separations, 
is $<0.5\%$ and does not reflect the larger relative populations 
of field stars of M--L spectral types (also see, for example, 
\citealt{politano04} and \citealt{dobbie05}).  
The \citet{farihi05a} fraction corresponds to $\lesssim1$ WD + brown 
dwarf binary expected in our sample of 154 red-excess WDs, which is, 
again, consistent with the presence of WD0145-221. 
Clearly, this result must be treated circumspectly because of 
the small number of expected, and observed, objects in this category, 
not to mention the fact that the likely range of sampled orbital 
separations is somewhat different between our survey and the 
Farihi surveys, which could produce an even smaller expected 
number of WD + brown dwarf binaries in our sample (because of the small
numbers of red-excess WDs and very small expected fraction of 
WD + brown dwarf binaries, we did not attempt to quantify or 
correct for this effect).  
However, \citet[][their Figure 6]{farihi05a} show that M5--9 stars 
{\em as binary companions to WDs} are found in only about 10\% of 
the observed systems, compared to a frequency of about 30\% in the 
field (however, as with the distribution of field stars, the peak 
spectral type for companions again occurs at M3--4).  
Thus, we should only expect $\approx15$ WD binaries containing 
an M5--9 star, which is more consistent with the observed (small) 
number of candidates whose nominal $(H-K_{\rm s})$ color is redward 
of that of a dM5 star.

\section{The Conclusions}

We have identified a large sample of candidates for binaries 
containing a WD and a cool main sequence star of spectral type 
as late as approximately M4--5.  However, we found only a small 
number of strong candidates for binaries containing companions 
of spectral types later than M5, and {\em no} new candidates for 
binaries containing an L type (i.e., potential brown dwarf) companion.  
This result is in agreement with the results from other recent 
searches for WD binaries, which suggest that the formation 
(and/or survival) rate for binaries with extreme mass 
ratios\footnote{That is, $M_2/M_{\rm pro}\lesssim0.05$, where 
$M_2$ is the initial mass of the companion and $M_{\rm pro}$ is 
the mass of the WD progenitor.  The initial companion mass could 
be smaller than the current companion mass due to material 
accreted by the companion during the post-main sequence evolution 
of the WD progenitor (summarized in \S5.4 of \citealt{farihi05a}).  
The progenitor mass, in turn, is likely a factor of $\gtrsim2$ 
larger than the current WD mass (up to $\sim10$ times larger for
massive WDs; e.g., \citealt{weidemann87,weidemann00}).} is very low.

An important result of our near-IR survey of WDs, begun in Paper I 
and culminating here, is simply the demonstration that the 
color-selection method is a viable technique for selecting binary 
candidates and has a high confirmed success rate.
The synthetic flux method (used in Paper III and \citealt{tremblay06}) 
is a logical {\em next} step, since it requires both additional data 
(e.g., multiple optical and infrared bands) and the calculation of 
synthetic photometry from model atmospheres for each candidate star 
(e.g., it would have been impractical to attempt this type of 
analysis as a first step for the entire MS99 catalog).
In the modern era of virtual observatories and all sky surveys 
(e.g., the upcoming all sky infrared surveys to be performed by the 
Wide-field Infrared Survey Explorer and Akari/ASTRO-F), the 
color-selection method that we have demonstrated using 2MASS 
photometry provides the most simple and efficient means to take 
advantage of existing, homogenous survey data in order to provide 
an initial list of binary candidates.

\acknowledgements

J.W.W. acknowledges financial support from a Harvey Mudd College 
National Merit Scholarship.  
M.G. acknowledges financial support from the Harvey S.\ Mudd Merit Award.
Thanks to Tom Marsh (University of Warwick) for making his 
collection of WD finding charts available, and 
to Jessica Hall (University of Southern California) and
Ryan Yamada (Harvey Mudd College) for their assistance in
identifying white dwarfs.
We also thank the anonymous referee for pointing out several papers 
that were helpful in improving the presentation of our results. 
This work was performed, in part, at the Jet Propulsion
Laboratory (JPL), California Institute of Technology (CIT).
Support for this work was provided by the National Aeronatuics
and Space Administration (NASA) under an Astrophysics Data Program  
grant issued through the Office of Space Science.
This research made use of the 
NASA/Infrared Processing and Analysis Center (IPAC) Infrared Science Archive, 
which is operated by JPL/CIT, under contract with NASA, and data 
products from the Two Micron All Sky Survey, which is a joint 
project of the University of Massachusetts and IPAC/CIT, funded 
by NASA and the National Science Foundation (NSF).  We utilized 
the SIMBAD database, operated at CDS, Strasbourg, France, and 
NASA's Astrophysics Data System.  The National Geographic 
Society-Palomar Observatory Sky Atlas (POSS-I) was made by CIT 
with grants from the National Geographic Society.  The Second 
Palomar Observatory Sky Survey (POSS-II) was made by CIT with 
funds from the NSF, the National Geographic Society, the Sloan 
Foundation, the Samuel Oschin Foundation, and the Eastman Kodak 
Corporation.

\appendix
\section{Notes on Individual Objects}
\label{s:t2notes}

\begin{description}
\item[WD0018-267:] Included (but not discussed) in \citet{kilic06}; 
does not show mid-IR excess compared to model WD in their Figure 1.

\item[WD0023+388:] Reference to binary status in MS99 is a 
private communication.

\item[WD0034-211:] MS99 note ``close double degenerate binary'' 
but \citet{bragaglia90} reclassified it as a WD+dM.

\item[WD0041+092:] BL Psc. 

\item[WD0116-231:] \citet{eggen78} and \citet{lamontagne00} classify 
WD0116-231 as DA+dM (the latter may just be repeating the 
Eggen \& Bessell classification); however, \citet{bessell79} list 
WD0116-231 in their table of composite spectrum stars (dK or dM plus 
a blue star) as type sd0.  Reference for binary status in MS99 is a 
preprint indicating the Montreal-Cambridge survey, with no obvious 
subsequent publication. 

\item[WD0130-196:] MS99 note ``MCT0130-1937 is a PG 1159 star with 
no detected variability.''

\item[WDJ0148-255:] WD0145-257.

\item[WD0232+035:] FS Cet, Feige 24.  Orbital period of 4.232 d 
\citep{thor78,RKcat}.

\item[WDJ0254-053:] WD0252-055, HD18131.

\item[WD0255+009.2:] The 2MASS colors do not strongly constrain a 
spectral type or luminosity class.

\item[WD0257+247:] Reference for binary status in MS99 is a private 
communication that probably refers to the wide ($a\sim140\arcsec$) 
common proper motion companion (NLTT 9583) reported by \citet{salim03}.

\item[WD0258+184:] Possibly an sdB star, with an unresolved cool (G8) 
companion, misclassified as a WD in MS99 \citep{lisker05}.

\item[WD0302+621:] \citet{schmidt95} find a marginal detection of an 
$\sim$kG magnetic field, but no sign of a red companion in its optical 
spectrum (they were not specifically looking for binaries, but noted 
the possible presence of a red spectral component in several other WDs 
in their survey).

\item[WD0308+096:] CC Cet.  Orbital period of 0.287 d \citep{RKcat}.

\item[WD0312+019:] The 2MASS colors do not strongly constrain a 
spectral type. 

\item[WDJ0357+286:] WD0353+284, V1092 Tau.  Orbital period 
of 0.365 d \citep{jeffries96a}. The WD may be accreting from the wind 
of its ultra-fast rotating companion \citep{jeffries96a,jeffries96b}.

\item[WD0413-077:] 40 Eri B.  The BC pair (A is a distant dK1 star) 
should be resolvable in 2MASS images but the dM star is either 
undetected or unseen due to the brightness of the WD.

\item[WD0416+272:] HL Tau 76, V411 Tau.  A ZZ Ceti variable (MS99).  
Not identified as a binary in \citet{farihi05a}.

\item[WD0419-487:] RR Cae.  Orbital period of 0.304 d \citep{RKcat}.

\item[WD0429+176:] HZ9.  Orbital period of 0.564 d \citep{RKcat}.

\item[WD0458-662:] Likely orbital period of $\sim0.7$ d to several 
days \citep{hutchings95}.

\item[WD0627+299:] Not identified as a binary in \citet{farihi05a}.

\item[WD0628-020:] Angular separation is near the 2MASS imaging 
resolution limit; however, no separate 2MASS data are available 
for the WD component, so the photometry results might be for the 
companion only.

\item[WD0710+741:] HR Cam.  Orbital period of 0.103 d \citep{RKcat}.

\item[WD0718-316:] IN CMa.  Orbital period of 1.262 d \citep{RKcat}.

\item[WD0800-533:] \citet{wick77} suggest that this object is a 
possible cataclysmic variable; they don't show its spectrum, but 
say that H$\alpha$ and H$\beta$ have emission line cores, and that 
it is a possible old nova.  Red near-IR colors reported independently 
by \cite{kawka07}.

\item[WD0802+387:] Hot DZ star with strong \ion{Ca}{2} H and K 
absorption lines \citep{sion90}.  Not identified as a binary in 
\citet{farihi05a}.

\item[WD0852+630:] Reference for binary status in MS99 refers to 
wide ($a\approx36\arcsec$) common proper motion companion.

\item[WD0858-220:] Reference for binary status in MS99 is a 
private communication.  Barely resolved in 2MASS images; only one 
source in 2MASS PSC, at position between the two stars, so the 
photometry could be for only the dM star or a blend of both.

\item[WD0908+226:] Reference for binary status in MS99 is a 
private communication (no other literature citations).

\item[WD0915+201:] Stellar image is slightly elongated 
(diameter $\approx3\arcsec$) in red POSS images, and shows a distinct 
color gradient from blue at the northernmost end to red at the 
southernmost end in the combined POSS B+R+IR 3-color image.  The 2MASS 
source position is offset toward the red end and there is no 
corresponding 2MASS source at the blue end, so it is possible that 
the 2MASS photometry corresponds only to the red star (i.e., the WD 
is undetected) or is a blend of both stars.  The 2MASS colors do 
not strongly constrain a spectral type.

\item[WD0937-095:] Reference for binary status in MS99 is a private 
communication, and probably refers to the $a\approx13\arcsec$ common 
proper motion companion.  There is some ambiguity in the literature 
about which star in the pair is the WD:\ 2MASS 09394969$-$0945562 is 
a closer match to the catalog coordinates of the WD \citep{salim03} 
than the southern component (2MASS 09394977$-$0946098); however, 
both stars are about equally red in th near-IR 
(e.g., $J-K_{\rm s}\approx0.80$--0.81).  We have used the northern 
component here, but the 2MASS photometry for the southern component 
was reported in Paper I.

\item[WD0949+451:] The red star is possibly a dM4.5+dM4.5 close 
binary separated by $0.009\arcsec$ (Paper III).  Reference for 
binary status in MS99 is a preprint citing the Hamburg-Schmidt 
Catalog with no published follow-up.

\item[WD1013-050:] Orbital period of 0.789 d \citep{RKcat}.

\item[WD1026+002:] UZ Sex.  Orbital period of 0.597 d \citep{RKcat}.

\item[WD1027-039:] Reference for binary status in MS99 is a private 
communication, and probably refers to the wide ($a\approx35\arcsec$) 
common proper motion companion reported by \citet{salim03}.

\item[WD1036-204:] MS99 note ``a polarized, carbon band, magnetic 
degenerate''; this might account for its near-IR colors instead of 
a red companion (see Paper II).

\item[WD1042-690:] Orbital period of 0.337 d \citep{RKcat}.

\item[WD1055-072:] An unusual DC type WD \citep{berg01}.  No evidence 
for an unresolved red companion in the surveys of 
\citet{debes05b,farihi05a,kilic06}.  Included (but not discussed) in 
\citet{kilic06}; does not show mid-IR excess compared to model WD in 
their Figure 1.

\item[WD1101+364:] MS99 note ``double degenerate DA+DA double-lined'' 
(also see \citealt{marsh95,nelemans05}).  Orbital period of 0.145 d 
\citep{RKcat}.

\item[WD1104+044:] Inspection of the POSS and 2MASS images shows that 
this is a not quite resolved ($a\approx3\arcsec$) common proper motion 
binary containing a very blue object (presumably the WD) and a red 
object; only the red object is detected by 2MASS.

\item[WD1106-211:] Although it is the closest star to the correct 
coordinates, it is possible that this is not the WD because it does 
not display the large proper motion ($\mu=0.467\arcsec$ y$^{-1}$) 
reported by \citet{evans92} (however, no star within several 
arcminutes shows such proper motion).

\item[WD1126+185:] Possibly an sdB+dG--K binary \citep{farihi05a}.

\item[WD1132-298:] ESO0439-095.  This spectroscopic binary has a pair 
of wide ($a\approx53\arcsec$) common proper motion companions 
(ESO0439-096A+B, which form an $a\approx4\arcsec$ binary themselves) 
that are dM2 and dM3 stars, respectively.

\item[WD1136+667:] Orbital period of 0.836 d \citep{RKcat}.

\item[WD1156+129:] This object has the correct optical brightness 
($V\sim17.5$), position, and proper motion ($\mu=0.05\arcsec$ y$^{-1}$ 
at position angle of $\theta=176^{\circ}$; \citealt{evans92}), 
but is very red ($V-J\approx+2.8$).

\item[WD1201+437:] \citet{xu99} classify this object as a quasar; 
however, their X-ray position error circle has a $9\arcsec$ radius.  
The extremely red 2MASS colors of this object 
(e.g., $H-K_{\rm s}\approx+1.1$) lend credence to reclassifying it 
as a quasar.

\item[WD1213+528:] EG UMa.  Orbital period of 0.668 d \citep{RKcat}.

\item[WD1224+309:] LM Com.  Orbital period of 0.259 d \citep{RKcat}.

\item[WD1247-176:] Orbital period of 0.571 d \citep{RKcat}.

\item[WDJ1255+258:] WD1253+261, IN Com.  Planetary nebula central 
star (PN G339.9+88.4).  The hot ($T>100,000$ K; \citealt{feibelman83}) 
component in IN Com is widely referred to as a subdwarf or WD precursor 
(e.g., \citealt{ritter86}).  It has been suggested in several 
literature sources that IN Com might be a triple system containing a 
close (possibly interacting) binary and a detached third component 
(likely a late M dwarf), although the identification of which stars 
are involved in the close vs.\ wide binaries, and their corresponding 
orbital periods, has not been conclusively established (summarized 
in \citealt{strassmeier97}).

\item[WD1305+018:] Stellar image is slightly elongated (diameter 
$\lesssim2\arcsec$) in the east-west direction in the POSS and 2MASS 
images; the west end is slightly bluer than the east end, but 
there is no strong color gradient.  Optical colors are redder than 
expected from its spectroscopic temperature, indicating a possible 
cool companion \citep{cheselka93}.

\item[WD1310-305:] Not a close double WD binary \citep{maxted99}.

\item[WD1314+293:] HZ43.  Reference for binary status in MS99 is a 
private communication.

\item[WD1330+793:] Reference for binary status in MS99 is a private 
communication.  Only the red, eastern component of this common proper 
motion binary is detected by 2MASS.

\item[WDJ1340+604:] WD1339+606.

\item[WD1412-109:] Stellar image appears slightly elongated in POSS 
and 2MASS images; there is a slight color gradient in the combined 
POSS B+R+IR 3-color image, with northwest end (faintly) red.

\item[WD1424+503:] There is a red (dMe) star $\approx7.5\arcsec$ north 
of the WD (e.g., \citealt{mason95}) that is easily resolved in the 
POSS and 2MASS images (and the 2MASS PSC); it is not known if this 
star is gravitationally bound with the WD.  In addition, there is 
previous evidence for the WD having a close (unresolved) red 
companion (summarized in \citealt{schwartz95}).

\item[WD1433+538:] Possibly a double degenerate \citep{liebert05}.

\item[WD1443+336:] Possible cataclysmic variable \citep{PG86}.

\item[WD1501+300:] This is the object closest to the (imprecise) 
target coordinates listed in MS99, and is identified in the University 
of Arizona White Dwarf Database finding chart.  However, there is a 
faint, blue star $\approx72\arcsec$ west that might be the actual WD.

\item[WD1517+502:] The companion in this binary is a dwarf carbon star; 
its 2MASS colors mimic very late M to early L spectral type.

\item[WD1541-381:] Reference for binary status in MS99 is a private 
communication, probably refers to the wide ($10.3\arcsec$) common 
proper motion companion reported by \citet{luyten49}.  This object 
is also reported as DA+dM \citep{bessell79}, but this might also 
refer to the wide companion.  The wide companion 
(2MASS 15451177$-$3818493) is slightly elongated in the 
northwest-southeast direction in the POSS (red) and 2MASS images 
and may, itself, be a close binary.  The 2MASS photometry for both 
stars (the alleged WD and the wide companion) is equally red: 
$(J-H)=+0.52$ and $(H-K_{\rm s})=+0.24$--$0.32$.

\item[WD1558+616:] Reference for binary status in MS99 is a preprint 
citing the Hamburg-Schmidt Catalog with no published follow-up.

\item[WD1603+125:] The WD has an 8 MG magnetic field \citep{wegner90a}.

\item[WD1608+118:] Stellar image is slightly elongated in the POSS images, 
with a strong color gradient in the combined POSS B+R+IR 3-color image 
from red at the northwest end to blue at the southeast end.  
The 2MASS detection is offset toward the red end.

\item[WD1610+383:] Inspection of POSS and 2MASS images shows that this 
is a common proper motion binary, barely resolved in the POSS images 
as a blue (southwest) and red (northeast) pair.  The single 2MASS 
detection is located closer to the red star -- see note in Paper I.  
There is also a much fainter star of intermediate (optical) color 
just north of the red component, but it does not appear to share 
proper motion with the binary.

\item[WD1619+525:] Stellar image is slightly elongated in east-west 
direction in (red) POSS images, with a color gradient in the combined 
POSS B+R+IR 3-color image from blue (east) to red (west).

\item[WD1622+323:] \citet{finley97} report an early M type companion 
from an optical spectrum of this object.

\item[WD1631+781:] Orbital period of 2.89 d \citep{RKcat}.

\item[WD1632-227.1:] Double degenerate (DC+DC) binary with separation 
of $a=5.75\arcsec$, resolved in all POSS and 2MASS images; 
component 1 is the northern, and somewhat brighter, of the two.  
We note that component 2 is even more red in the near-IR 
($J-H\approx+0.7$, $H-K_{\rm s}\approx+0.5$), but is not included here 
because it is a Moderate 2MASS All Sky detection 
($\sigma_{K_{\rm s}}>0.1$ mag).  The 2MASS colors do not strongly 
constrain a spectral type.

\item[WD1634-573:] MS99 note ``Hot (55,000 K) DO degenerate with 
photospheric carbon in UV.''

\item[WD1643+143:] Optical photometry is too red for an isolated DA WD, 
suggesting an unresolved cool companion \citep{kidder91}.

\item[WD1654+160:] V824 Her.  Pulsating WD \citep{winget84}.

\item[WDJ1711+667:] WD1711+668.  The 2MASS colors do not strongly 
constrain a spectral type.

\item[WD1717-345:] The 2MASS colors do not strongly constrain a 
spectral type.

\item[WDJ1820+580:] WD1819+580.

\item[WD1845+019:] Spectroscopic evidence for a long-period, 
red companion inferred from H$\alpha$ emission line radial 
velocities \citep{maxted99}.  It is not known if the close 
neighbor star described by \citet{debes05b} is gravitationally 
bound to the WD and if it is the same spectroscopic companion 
star inferred by \citet{maxted99}.  \citet{schmidt95} find no 
sign of a red companion in its optical spectrum (they were 
not specifically looking for binaries, but noted the possible 
presence of a red spectral component in several other WDs in 
their survey for magnetic fields).

\item[WD2009+622:] Orbital period of 0.741 d \citep{RKcat}.

\item[WDJ2013+400:] WD2011+395.  Orbital period of 0.706 d \citep{RKcat}.

\item[WDJ2024+200:] WD2022+198.

\item[WD2108-431:] This object is the one indicated in the University 
of Arizona White Dwarf Database finding chart, but it is not blue in 
optical (POSS) images.  There is a blue object $\approx34\arcsec$ east, 
but that object is fuzzy and probably a galaxy.

\item[WD2131+066:] IR Peg.  Orbital period of 0.164 d \citep{RKcat}.  
Pulsating (PG1159 type) WD (MS99).  The 2MASS colors do not strongly 
constrain a spectral type.  \citet{paunzen98} suggest that the star 
observed at $0.3\arcsec$ separation is not the binary companion but, 
rather, the companion is an M dwarf in an even closer, as yet 
unresolved, orbit.

\item[WD2133+463:] Common proper motion ($\mu=0.46\arcsec$ y$^{-1}$) 
binary with separation of $a\approx3\arcsec$ \citep{bakos02}.  
Possibly only the red component is detected by 2MASS (e.g., the stellar 
image doesn't appear elongated in 2MASS images).

\item[WD2154+408:] Orbital period of 0.268 d (\citealt{RKcat}; 
also see \citealt{hilwig02}).

\item[WD2237+819:] Orbital period of 0.124 d \citep{RKcat}.  Reference 
for binary status in MS99 is a preprint indicating the Hamburg-Schmidt 
Survey (e.g., see \citealt{boris04}).

\item[WD2256+249:] MS Peg.  Orbital period of 0.174 d \citep{RKcat}.

\item[WD2311-068:] Not identified as a binary in \citet{farihi05a}.  
This is a cool DQ6 type WD (e.g., \citealt{dufour05}), which might 
account for its red near-IR colors.

\item[WD2318-137:] This is a known common proper motion binary 
(MS99, \citealt{silvestri02}); there is a co-moving blue object 
(a WD?) located $\approx89\arcsec$ east of the red star whose 
2MASS designation is listed here -- this is likely the companion 
noted in MS99 and \citep{silvestri02}.  However, the red star 
appears elongated in the southeast-to-northwest direction on the 
POSS (red) images, suggesting that it might also be a binary 
(with separation of $a\approx3\arcsec$).  The POSS blue and 2MASS 
images show the least elongation, suggesting that this closer 
binary has red and blue components (and possibly only the red 
component is detected by 2MASS).

\item[WD2326-224:] Stellar image is slightly elongated in all POSS 
images and the combined POSS B+R+IR three-color image shows a color 
gradient from blue (northwest) to red (southeast).  The stellar image is 
not noticeably elongated in the 2MASS images, and the coordinates of 
the single 2MASS detection are offset toward the red end of the 
elongated optical image.

\item[WD2336-187:] Believed to be a single WD \citep{tremblay06}. 

\end{description}


\begin{deluxetable}{llllllllllll}
\tabletypesize{\scriptsize}
\tablewidth{0pt}
\tablecaption{2MASS Photometry for White Dwarfs with Near-IR Excess
\label{t:phot}}
\tablehead{
\colhead{WD}  & \colhead{2MASS} & \colhead{$J$} & \colhead{$H$} & \colhead{$K_{\rm s}$} 
}
\startdata
0014+097     & 00165618+1003591   & 12.414(25) & 11.880(29) & 11.602(22) \\
0017+061     & 00194100+0624075   & 13.738(33) & 13.189(34) & 12.983(32) \\
0018$-$267   & 00213073$-$2626114 & 12.504(26) & 12.106(25) & 12.006(23) \\
0023+388     & 00263309+3909044   & 13.810(26) & 13.268(30) & 12.939(33) \\
0027$-$549   & 00294996$-$5441354 & \phn9.758(24) & \phn9.129(23) & \phn8.883(23) \\
0034$-$211   & 00372502$-$2053422 & 11.454(23) & 10.884(21) & 10.648(26) \\
0041+092     & 00440131+0932578   & \phn8.450(34) & \phn7.927(34) & \phn7.805(29) \\
0104$-$331   & 01064686$-$3253124 & 14.740(38) & 14.162(44) & 13.913(57) \\
0116$-$231   & 01183718$-$2254578 & 14.640(42) & 14.100(37) & 13.828(55) \\
0130$-$196   & 01323935$-$1921394 & 14.749(37) & 14.253(32) & 13.997(57) \\
0131$-$163   & 01342407$-$1607083 & 12.966(27) & 12.468(28) & 12.215(30) \\
0145$-$705   & 01461179$-$7020197 & 14.962(41) & 14.538(63) & 14.383(77) \\
0145$-$221   & 01472183$-$2156512 & 14.923(32) & 14.450(45) & 14.335(64) \\
0148$-$255J  & 01480822$-$2532452 & 12.412(26) & 11.830(21) & 11.594(23) \\
0205+133     & 02080350+1336256   & 12.799(22) & 12.198(24) & 11.961(20) \\
0208$-$153   & 02104280$-$1506356 & 12.589(24) & 12.057(25) & 11.772(25) \\
0232+035     & 02350758+0343567   & 11.265(24) & 10.733(21) & 10.557(19) \\
0237+115     & 02400663+1148280   & 13.788(30) & 13.106(27) & 12.892(32) \\
0248+601     & 02520803+6019428   & 13.529(21) & 12.894(28) & 12.673(29) \\
0254$-$053J  & 02543883$-$0519509 & \phn5.709(19) & \phn5.263(47) & \phn5.090(18) \\
0255+009.2   & 02581788+0109458   & 15.560(63) & 14.824(69) & 14.526(83) \\
0257+247     & 03003606+2453393   & 12.670(19) & 12.070(16) & 11.851(21) \\
0258+184     & 03011287+1840539   & 14.991(40) & 14.803(67) & 14.475(76) \\
0303$-$007   & 03060719$-$0031144 & 13.164(24) & 12.627(27) & 12.405(26) \\
0302+621     & 03061669+6222226   & 15.015(44) & 14.989(89) & 14.749(95) \\
0308+096     & 03105499+0949256   & 13.723(29) & 13.183(31) & 12.934(35) \\
0309$-$275   & 03113318$-$2719260 & 13.514(26) & 12.885(34) & 12.731(30) \\
0312+019     & 03145212+0206072   & 15.603(51) & 14.998(55) & 14.776(98) \\
0347$-$137   & 03501451$-$1335138 & 12.080(29) & 11.540(29) & 11.296(23) \\
0357+286J    & 03570582+2837516   & \phn9.843(23) & \phn9.275(24) & \phn9.057(17) \\
0354+463     & 03581711+4628397   & 13.594(27) & 13.084(38) & 12.727(27) \\
0357$-$233   & 03590491$-$2312243 & 14.961(44) & 14.594(60) & 14.256(65) \\
0357+081     & 04002668+0814069   & 14.562(38) & 14.343(56) & 14.122(57) \\
0413$-$077   & 04152173$-$0739173 & \phn6.747(20) & \phn6.278(40) & \phn5.962(26) \\
0416+272     & 04185663+2717484   & 15.134(38) & 15.361(90) & 15.065(94) \\
0419$-$487   & 04210556$-$4839070 & 10.720(24) & 10.148(23) & \phn9.852(25) \\
0429+176     & 04322373+1745026   & 10.753(21) & 10.161(19) & \phn9.913(17) \\
0430+136     & 04331053+1345134   & 13.533(21) & 12.877(23) & 12.634(26) \\
0458$-$662   & 04585395$-$6628134 & 13.437(26) & 12.727(27) & 12.513(29) \\
0627+299     & 06303679+2956180   & 15.128(38) & 15.176(71) & 14.924(89) \\
0628$-$020   & 06303881$-$0205537 & 10.729(27) & 10.144(26) & \phn9.857(24) \\
0710+741     & 07170975+7400406   & 14.692(33) & 14.423(61) & 14.148(65) \\
0718$-$316   & 07204790$-$3147027 & 13.253(25) & 12.749(26) & 12.502(27) \\
0752$-$146   & 07550896$-$1445506 & 12.621(24) & 12.141(23) & 11.837(19) \\
0800$-$533   & 08020022$-$5327499 & 13.703(33) & 13.186(27) & 12.991(27) \\
0802+387     & 08055764+3833444   & 15.336(47) & 15.193(79) & 14.899(91) \\
0805+654     & 08094545+6518172   & 14.002(24) & 13.449(27) & 13.171(31) \\
0812+478     & 08154894+4740391   & 14.587(32) & 14.165(41) & 13.882(47) \\
0824+288     & 08270508+2844024   & 12.423(28) & 11.802(28) & 11.650(30) \\
0852+630     & 08562826+6250226   & 12.827(24) & 12.122(33) & 11.953(21) \\
0858$-$220   & 09005722$-$2213505 & 11.348(27) & 10.771(24) & 10.572(21) \\
0908+226     & 09114308+2227488   & 15.160(40) & 14.492(35) & 14.389(59) \\
0915+201     & 09183291+1953070   & 15.721(58) & 15.166(78) & 14.867(80) \\
0928+399     & 09315566+3946076   & 15.049(41) & 14.453(58) & 14.100(56) \\
0933+025     & 09354067+0222005   & 13.270(24) & 12.726(23) & 12.477(24) \\
0937$-$095   & 09394969$-$0945562 & 14.715(33) & 14.135(38) & 13.898(48) \\
0949+451     & 09522209+4454288   & 11.924(22) & 11.335(16) & 11.035(13) \\
0950+185     & 09524582+1821026   & 12.691(22) & 12.011(22) & 11.786(20) \\
0956+045     & 09583717+0421292   & 14.659(40) & 14.131(46) & 13.837(59) \\
1001+203     & 10040431+2009226   & 12.640(21) & 12.028(21) & 11.766(20) \\
1004$-$178   & 10070776$-$1805246 & 12.555(22) & 11.983(27) & 11.745(23) \\
1013$-$050   & 10162867$-$0520320 & 10.607(27) & \phn9.990(25) & \phn9.770(23) \\
1015$-$173   & 10172883$-$1737059 & 15.238(50) & 14.874(58) & 14.552(99) \\
1026+002     & 10283487$-$0000295 & 11.751(24) & 11.219(27) & 10.943(21) \\
1027$-$039   & 10295925$-$0413014 & 15.207(50) & 14.611(38) & 14.286(71) \\
1033+464     & 10362522+4608312   & 12.564(22) & 12.032(24) & 11.752(18) \\
1036$-$204   & 10385559$-$2040572 & 14.633(33) & 14.346(41) & 14.035(67) \\
1037+512     & 10401680+5056468   & 13.796(24) & 13.261(26) & 12.972(26) \\
1042$-$690   & 10441023$-$6918180 & 11.423(26) & 10.896(27) & 10.561(21) \\
1049+103     & 10522772+1003380   & 13.266(29) & 12.830(35) & 12.483(33) \\
1055$-$072   & 10573517$-$0731233 & 13.770(29) & 13.680(32) & 13.485(38) \\
1101+364     & 11043257+3610490   & 14.832(38) & 14.962(70) & 14.714(95) \\
1104+044     & 11070472+0409077   & 14.127(33) & 13.674(35) & 13.509(53) \\
1106+316     & 11084306+3123559   & 15.136(41) & 14.524(44) & 14.429(86) \\
1106$-$211   & 11091095$-$2123320 & 14.676(33) & 13.931(39) & 13.817(57) \\
1123+189     & 11261906+1839178   & 12.754(23) & 12.217(19) & 11.990(20) \\
1126+185     & 11291804+1816457   & 12.643(23) & 12.166(31) & 12.094(23) \\
1132$-$298   & 11352186$-$3010156 & 13.799(32) & 13.164(29) & 12.964(35) \\
1133+358     & 11354300+3534235   & 11.625(19) & 11.079(21) & 10.802(19) \\
1136+667     & 11390593+6630184   & 12.314(25) & 11.695(26) & 11.543(23) \\
1141+504     & 11434997+5010203   & 14.779(40) & 14.215(43) & 13.989(48) \\
1147+371     & 11503125+3654159   & 14.892(39) & 14.271(48) & 14.114(48) \\
1156+129     & 11591564+1239299   & 14.721(32) & 14.099(30) & 13.895(42) \\
1201+437     & 12042403+4330570   & 15.393(51) & 14.857(68) & 13.785(41) \\
1210+464     & 12125961+4609467   & 12.035(23) & 11.396(21) & 11.161(20) \\
1213+528     & 12154411+5231013   & \phn9.979(22) & \phn9.340(27) & \phn9.033(21) \\
1214+032     & 12165190+0258046   & \phn9.234(18) & \phn8.671(22) & \phn8.422(18) \\
1218+497     & 12210535+4927207   & 14.588(38) & 14.002(36) & 13.837(60) \\
1224+309     & 12263089+3038527   & 15.129(48) & 14.669(68) & 14.393(77) \\
1246+299     & 12484722+2942507   & 15.055(43) & 14.479(55) & 14.373(79) \\
1247$-$176   & 12502208$-$1754465 & 13.502(24) & 12.863(23) & 12.601(31) \\
1255+258J    & 12553374+2553308   & \phn7.370(27) & \phn6.947(51) & \phn6.855(24) \\
1254$-$133   & 12563956$-$1334421 & 14.499(33) & 13.973(51) & 13.706(51) \\
1305+018     & 13075472+0132106   & 12.982(26) & 12.400(23) & 12.144(21) \\
1307$-$141   & 13102255$-$1427099 & 13.849(23) & 13.226(35) & 13.013(36) \\
1310$-$305   & 13134158$-$3051336 & 14.931(35) & 15.125(53) & 14.915(95) \\
1314+293     & 13162169+2905548   & 10.373(19) & \phn9.807(29) & \phn9.565(25) \\
1319$-$288   & 13224044$-$2905347 & 12.754(24) & 12.271(23) & 11.986(23) \\
1330+793     & 13303294+7905126   & 12.506(21) & 11.880(23) & 11.657(19) \\
1333+487     & 13360209+4828472   & 11.794(23) & 11.200(17) & 10.927(23) \\
1340+604J    & 13410002+6026104   & 15.555(58) & 15.143(74) & 14.776(82) \\
1401+005     & 14034531+0021359   & 13.443(23) & 12.851(26) & 12.771(30) \\
1412$-$049   & 14150220$-$0511040 & 13.803(30) & 13.092(27) & 12.988(34) \\
1412$-$109   & 14150761$-$1109213 & 15.379(64) & 14.588(66) & 14.391(73) \\
1415+132     & 14174025+1301487   & 14.263(36) & 13.725(46) & 13.553(46) \\
1424+503     & 14264410+5006274   & 10.626(20) & 10.008(15) & \phn9.822(22) \\
1433+538     & 14344329+5335212   & 14.667(35) & 14.216(50) & 13.919(45) \\
1435+370     & 14373667+3651378   & 13.457(24) & 12.965(25) & 12.746(28) \\
1436$-$216   & 14391264$-$2150138 & 13.312(27) & 12.770(22) & 12.519(29) \\
1443+336     & 14460066+3328502   & 14.284(30) & 13.725(30) & 13.516(40) \\
1458+171     & 15001934+1659146   & 14.701(31) & 14.209(45) & 13.847(47) \\
1501+300     & 15032075+2952584   & 14.178(27) & 13.745(26) & 13.767(53) \\
1502+349     & 15043185+3446584   & 15.231(45) & 14.766(61) & 14.314(67) \\
1504+546     & 15060542+5428186   & 13.847(25) & 13.260(26) & 13.001(27) \\
1517+502     & 15190599+5007027   & 15.559(60) & 14.746(71) & 14.157(72) \\
1522+508     & 15242519+5040098   & 14.757(35) & 14.141(45) & 13.959(58) \\
1541$-$381   & 15451092$-$3818515 & 11.953(25) & 11.432(28) & 11.190(25) \\
1558+616     & 15585553+6132037   & 14.204(32) & 13.600(42) & 13.368(42) \\
1603+125     & 16053211+1225429   & 13.552(23) & 13.120(23) & 12.977(26) \\
1608+118     & 16105319+1143538   & 12.059(22) & 11.482(23) & 11.260(19) \\
1610+383     & 16122147+3812299   & 14.437(34) & 13.807(36) & 13.521(42) \\
1619+525     & 16202428+5223215   & 14.168(32) & 13.545(35) & 13.425(42) \\
1619+414     & 16211268+4118093   & 13.937(21) & 13.311(29) & 13.025(27) \\
1622+323     & 16244899+3217021   & 14.633(29) & 13.963(31) & 13.773(39) \\
1631+781     & 16291031+7804399   & 10.975(21) & 10.398(21) & 10.164(14) \\
1632$-$227.1 & 16352428$-$2250279 & 15.279(56) & 14.654(75) & 14.395(92) \\
1634$-$573   & 16383108$-$5728112 & \phn7.120(23) & \phn6.699(44) & \phn6.571(29) \\
1643+143     & 16453913+1417462   & 12.732(24) & 12.125(31) & 11.957(24) \\
1646+062     & 16490776+0608453   & 14.035(27) & 13.424(30) & 13.237(35) \\
1654+160     & 16565765+1556254   & 13.059(29) & 12.421(33) & 12.129(24) \\
1711+667J    & 17112736+6645319   & 15.120(43) & 14.457(57) & 14.211(87) \\
1717$-$345   & 17211032$-$3433286 & 12.870(39) & 12.208(60) & 11.940(54) \\
1820+580J    & 18202977+5804410   & 14.058(32) & 13.718(35) & 13.401(39) \\
1833+644     & 18332921+6431520   & 14.098(30) & 13.516(30) & 13.241(38) \\
1845+019     & 18473908+0157356   & 12.398(53) & 12.014(82) & 11.520(47) \\
1844$-$654   & 18490202$-$6525144 & 12.703(24) & 12.045(26) & 11.833(21) \\
1950+279     & 19522838+2807527   & 15.187(48) & 14.635(60) & 14.527(80) \\
2009+622     & 20104287+6225321   & 14.277(31) & 13.915(31) & 13.578(48) \\
2013+400J    & 20130936+4002242   & 13.044(24) & 12.520(24) & 12.260(32) \\
2024+200J    & 20241609+2000474   & 14.370(33) & 13.845(43) & 13.595(47) \\
2101$-$364   & 21044692$-$3615251 & 14.515(26) & 13.915(32) & 13.756(38) \\
2108$-$431   & 21113740$-$4258116 & 12.773(24) & 12.161(23) & 11.934(25) \\
2118$-$333   & 21214687$-$3310477 & 12.595(24) & 11.950(25) & 11.734(25) \\
2131+066     & 21340822+0650573   & 15.317(42) & 14.723(60) & 14.486(92) \\
2133+463     & 21351760+4633174   & 11.297(21) & 10.756(17) & 10.460(19) \\
2151$-$015   & 21540644$-$0117102 & 12.452(29) & 11.778(22) & 11.414(27) \\
2154+408     & 21561824+4102452   & 12.877(25) & 12.383(33) & 12.148(27) \\
2237+819     & 22371556+8210273   & 12.824(26) & 12.282(32) & 11.978(24) \\
2256+249     & 22584811+2515439   & 11.675(20) & 11.180(25) & 10.915(18) \\
2311$-$068   & 23142520$-$0632475 & 14.951(36) & 14.942(71) & 14.730(93) \\
2317+268     & 23200401+2706237   & 14.609(33) & 14.074(36) & 13.783(50) \\
2318$-$137   & 23210825$-$1327465 & 12.074(22) & 11.444(24) & 11.174(21) \\
2326$-$224   & 23283886$-$2210209 & 12.627(26) & 12.035(23) & 11.766(21) \\
2336$-$187   & 23385279$-$1826123 & 15.057(40) & 14.939(63) & 14.681(93) 
\enddata
\end{deluxetable}

\begin{deluxetable}{llllllllllll}
\tabletypesize{\scriptsize}
\tablewidth{0pt}
\tablecaption{Binary Parameters
\label{t:binary}}
\tablehead{
\colhead{WD}  & \colhead{Binary} & \colhead{References} & \colhead{Spectral} & \colhead{$a_{\rm sep}$}  & \colhead{Data} & \colhead{References} & \colhead{Comments?} \\
\colhead{ }   & \colhead{Status} & \colhead{ }          & \colhead{Types}          & \colhead{($\arcsec$)}    & \colhead{ }    & \colhead{ }          & \colhead{ }         \\
\colhead{(1)} & \colhead{(2)}    & \colhead{(3)}        & \colhead{(4)}        & \colhead{(5)}            & \colhead{(6)}  & \colhead{(7)}        & \colhead{(8)}        
}
\startdata
0014+097   & confirmed & 36 & WD+dM  & \nodata     & S  & 36        & \nodata \\
           &           &    & dM2--4 & $\lesssim4$ & IP & this work & \\[6pt]

0017+061   & confirmed & 7, 8 & DA2+dM4  & 2.0         & IP & 8         & \nodata \\
           &           &      & dM0--2.5 & $\lesssim4$ & IP & this work & \\[6pt]

0018-267   & tentative & this work & dG8--K2 & $\lesssim4$ & IP & this work & see Appendix \ref{s:t2notes} \\[6pt]

0023+388   & confirmed & 43, Papers I \& III & WD+dM5.5 & $\lesssim0.025$ & IP & Paper III & see Appendix \ref{s:t2notes} \\
           &           &                     & dM3--5   & $\lesssim4$     & IP & this work & \\[6pt]

0027-549   & confirmed & 47, 62 & DA+dM3   & 4           & IS & 47, 62    & \nodata \\
           &           &        & dM1--2.5 & $\lesssim4$ & IP & this work & \\[6pt]

0034-211   & confirmed & 8, 43, Papers I \& III & WD+dM3.5 & 0.328       & IP & Paper III & see Appendix \ref{s:t2notes} \\
           &           &                        & DA3+dM3  & $\leq0.5$   & IP & 8         & \\
           &           &                        & dM1--3   & $\lesssim4$ & IP & this work & \\[6pt]

0041+092   & confirmed & 30, 43 & DA+dK2    & $\lesssim0.08$ & IP  & 2         & see Appendix \ref{s:t2notes} \\
           &           &        & DA+dK1--3 & $<1.9$         & IPS & 30        & \\
           &           &        & dK3--5    & $\lesssim4$    & IP  & this work & \\[6pt]

0104-331   & candidate & this work & dM0--4.5 & $\lesssim4$ & IP & this work & \nodata \\[6pt]

0116-231   & confirmed & Papers I \& III & WD+dM4.5 & 1.105       & IP & Paper III & see Appendix \ref{s:t2notes} \\
           &           &                 & dM1--5   & $\lesssim4$ & IP & this work & \\[6pt]

0130-196   & candidate & Paper I & dM0--4.5 & $\lesssim4$ & IP & this work & see Appendix \ref{s:t2notes} \\[6pt]

0131-163   & confirmed & 8, Papers I \& III & WD+dM3.5 & 0.189       & IP & Paper III & \nodata \\
           &           &                    & DA1+dM2  & $\leq0.5$   & IP & 8         & \\
           &           &                    & dM1--3   & $\lesssim4$ & IP & this work & \\[6pt]

0145-705   & candidate & this work & dG6--dK3 & $\lesssim4$ & IP & this work & \nodata \\[6pt]

0145-221   & confirmed & 8, Paper I & DA4+dL6 & $\leq0.3$   & IP & 8         & \nodata \\
           &           &            & dK1--4  & $\lesssim4$ & IP & this work & \\[6pt]

J0148-255  & confirmed & 14, 43, 45, 79, Papers I \& III & WD+dM3.5  & 2.295        & IP & Paper III & see Appendix \ref{s:t2notes} \\
           &           &                                 & dM1--3    & $\lesssim4$  & IP & this work & \\
           &           &                                 & WD+dM3--4 & $\lesssim10$ & IP & 14        & \\[6pt]

0205+133   & confirmed & 15, Papers I \& III & WD+dM1 & 1.257       & IP & Paper III & \nodata \\
           &           &                     & dM0--3 & $\lesssim4$ & IP & this work & \\[6pt]

0208-153   & confirmed & Papers I \& III & WD+dM2     & 2.647       & IP & Paper III & \nodata \\
           &           &                 & dM2.5--4.5 & $\lesssim4$ & IP & this work & \\[6pt]

0232+035   & confirmed & 43, 66 & DA+dM   & close       & S  & 18, 66    & see Appendix \ref{s:t2notes} \\
           &           &        & dK3--M1 & $\lesssim4$ & IP & this work & \\[6pt]

0237+115   & confirmed & 15, 43, Paper III & WD+dM3 & 0.124       & IP & Paper III & \nodata \\
           &           &                   & dM0--2 & $\lesssim4$ & IP & this work &  \\[6pt]

0248+601   & candidate & this work & dM0--2 & $\lesssim4$ & IP & this work & \nodata \\[6pt]

J0254-053  & confirmed & 43, 69 & DA+K0 IV            & \nodata        & S  & 69        & see Appendix \ref{s:t2notes} \\
           &           &        & DA+K0 IV--III       & $\lesssim0.08$ & IP & 2         & \\
           &           &        & dK0--4 (G5--K0 III) & $\lesssim4$    & IP & this work & \\[6pt]

0255+009.2 & confirmed & 51, 60 & DA+dM3e     & close       & S  & 60        & see Appendix \ref{s:t2notes} \\
           &           &        & dM or M III & $\lesssim4$ & IP & this work &  \\[6pt]

0257+247   & candidate & this work & dM0--2.5 & $\lesssim4$ & IP & this work & see Appendix \ref{s:t2notes} \\[6pt]

0258+184   & tentative & 38, this work & \nodata & $\lesssim4$ & IP & this work & see Appendix \ref{s:t2notes} \\[6pt]

0303-007   & confirmed & 43, Papers I \& III, 60 & DA+dM2 & close           & S  & 75, 60    & \nodata \\
           &           &                         & WD+dM4 & $\lesssim0.025$ & IP & Paper III & \\
           &           &                         & dM1--3 & $\lesssim4$     & IP & this work & \\[6pt]

0302+621   & tentative & this work & \nodata & $\lesssim4$ & IP & this work & see Appendix \ref{s:t2notes} \\[6pt]

0308+096   & confirmed & 8, 15, 43, 54 & DA2+dM4.5 & close       & IP & 8         & see Appendix \ref{s:t2notes} \\
           &           &               & dM1--4    & $\lesssim4$ & IP & this work & \\[6pt]

0309-275   & candidate & Paper I & dK7--M0 & $\lesssim4$ & IP & this work & \nodata \\[6pt]

0312+019   & candidate & this work & dK5--M5 & $\lesssim4$ & IP & this work & see Appendix \ref{s:t2notes} \\[6pt]

0347-137   & confirmed & 8, Papers I \& III & WD+dM4.5 & 1.052       & IP & Paper III  & \nodata \\
           &           &                    & DA2+dM3  & $\leq0.5$   & IP & 8          & \\
           &           &                    & dM1--3   & $\lesssim4$ & IP & thiss work & \\[6pt]

J0357+286  & confirmed & 26, 27, 43, Paper I & WD+dK2   & close       & S  & 26, 27    & see Appendix \ref{s:t2notes} \\
           &           &                     & dK4--dM2 & $\lesssim4$ & IP & this work & \\[6pt]

0354+463   & confirmed & 8, 43, Paper III & WD+dM7   & $\lesssim0.025$ & IP & Paper III & \nodata \\
           &           &                  & DA6+dM7  & $\leq0.5$       & IP & 8         & \\
           &           &                  & dM4--5.5 & $\lesssim4$     & IP & this work & \\[6pt]

0357-233   & confirmed & 8, Papers I \& III & WD+dM3  & 1.190       & IP & Paper III & \nodata \\
           &           &                    & DA1+dM3 & 1.2         & IP & 8         & \\
           &           &                    & \nodata & $\lesssim4$ & IP & this work & \\[6pt]

0357+081   & tentative & this work & \nodata & $\lesssim4$ & IP & this work & \nodata \\[6pt]

0413-077   & confirmed & 43, Paper I & WD+dM4.5e & $\approx7$ & I & 68, 9 & see Appendix \ref{s:t2notes} \\[6pt]

0416+272   & tentative & this work & \nodata & $\lesssim4$ & IP & this work & see Appendix \ref{s:t2notes} \\[6pt]

0419-487   & confirmed & 3, 42 & WD+dM6   & close       & SP & 3         & see Appendix \ref{s:t2notes} \\
           &           &       & WD+dM4   & close       & S  & 42        & \\
           &           &       & dM3--4.5 & $\lesssim4$ & IP & this work & \\[6pt]

0429+176   & confirmed & 35, 43, Paper I & DA+dM4.5e & close       & SP & 35        & see Appendix \ref{s:t2notes} \\
           &           &                 & dM2--3    & $\lesssim4$ & IP & this work & \\[6pt]

0430+136   & confirmed & 43, 75, Paper I & DA+dM      & \nodata     & S  & 75        & \nodata \\
           &           &                 & dM0.5--2.5 & $\lesssim4$ & IP & this work & \\[6pt]

0458-662   & confirmed & 24, 43, Papers I \& III & DA+dM2   & close           & PS & 24        & see Appendix \ref{s:t2notes} \\
           &           &                         & WD+dM2.5 & $\lesssim0.025$ & IP & Paper III & \\[6pt]

0627+299   & tentative & this work & \nodata & $\lesssim4$ & IP & this work & see Appendix \ref{s:t2notes} \\

0628-020   & confirmed & 43, 47, 62, Paper I & WD+dM      & 4           & IS & 47        & see Appendix \ref{s:t2notes} \\
           &           &                     & dM2.5--4.5 & $\lesssim4$ & IP & this work & \\[6pt]

0710+741   & confirmed & 8, Paper I & DA3+dM7 & close & IP & 8 & see Appendix \ref{s:t2notes} \\[6pt]

0718-316   & confirmed & 43, 70 & DAO+dM0--2 & close & S & 70 & see Appendix \ref{s:t2notes} \\[6pt]

0752-146   & confirmed & 8, Paper I & DA3+dM6 & close & IP & 8 & \nodata \\[6pt]

0800-533   & candidate & 77, 29, this work & dM3--4           & \nodata     & P  & 29        & see Appendix \ref{s:t2notes} \\
           &           &                   & dK3--5 or dM2--3 & $\lesssim4$ & IP & this work & \\[6pt]

0802+387   & tentative & this work & \nodata & $\lesssim4$ & IP & this work & see Appendix \ref{s:t2notes} \\[6pt]

0805+654   & confirmed & 36 & WD+dM  & \nodata     & S  & 36        & \nodata \\
           &           &    & dM2--4 & $\lesssim4$ & IP & this work & \\[6pt]

0812+478   & candidate & Paper I & \nodata & $\lesssim4$ & IP & this work & \nodata \\[6pt]

0824+288   & confirmed & 8, 15, 36, 43          & DA1+dC+dM3.5 & $\leq0.5$, 3.3 & IP & 8         & \nodata \\
           &           &                        & dK6--M1      & $\lesssim4$    & IP & this work & \\[6pt]

0852+630   & candidate & this work & \nodata & $\lesssim4$ & IP & this work & see Appendix \ref{s:t2notes} \\[6pt]

0858-220   & confirmed & 43, 55, 61 & DC+dM   & 5       & IPS & 55        & see Appendix \ref{s:t2notes} \\
           &           &            & dK5--M2 & $\sim4$ & IP  & this work & \\[6pt]

0908+226   & candidate & 43, Paper I & late dK & $\lesssim4$ & IP & this work & see Appendix \ref{s:t2notes} \\[6pt]

0915+201   & confirmed & Paper I, this work & mid dK -- late dM & $\lesssim4$ & IP & this work & see Appendix \ref{s:t2notes} \\[6pt]

0928+399   & confirmed & 60 & DA+dM3e & close       & S  & 60        & \nodata \\
           &           &    & dM3--8  & $\lesssim4$ & IP & this work & \\[6pt]

0933+025   & confirmed & 8, 15, 36, 43 & DA2+dM3.5 & $\leq0.5$   & IP & 8         & \nodata \\
           &           &               & dM2--3    & $\lesssim4$ & IP & this work & \\[6pt]

0937-095   & candidate & Paper I & dK4--M4 & $\lesssim4$ & IP & this work & see Appendix \ref{s:t2notes} \\[6pt]

0949+451   & confirmed & Paper III & WD+dM4.5 & 2.892       & IP & Paper III & see Appendix \ref{s:t2notes} \\
           &           &           & dM3--4.5 & $\lesssim4$ & IP & this work & \\[6pt]

0950+185   & confirmed & 8, 15, 36, 43 & DA2+dM2 & 1.1         & IP & 8         & \nodata \\
           &           &               & dM0--1  & $\lesssim4$ & IP & this work & \\[6pt]

0956+045   & confirmed & 8, 15, 43 & DA3+dM4.5 & 2.0         & IP & 8         & \nodata \\
           &           &           & dM2--5.5  & $\lesssim4$ & IP & this work & \\[6pt]

1001+203   & confirmed & 8, 15, 36, 43, Paper I & DA2+dM3  & $\leq0.5$   & IP & 8         & \nodata \\
           &           &                        & dM1--3.5 & $\lesssim4$ & IP & this work & \\[6pt]

1004-178   & candidate & this work & dM0.5--3 & $\lesssim4$ & IP & this work & \nodata \\[6pt]

1013-050   & confirmed & 8, 43, Paper I & DAO1+dM4.5+dM1+dM1 & close, 3.2, 3.2 & IP & 8         & see Appendix \ref{s:t2notes} \\
           &           &                & dM0--2.5           & $\lesssim4$     & IP & this work & \\[6pt]

1015-173   & candidate & Paper I & \nodata & $\lesssim4$ & IP & this work & \nodata \\[6pt]

1026+002   & confirmed & 8, 15, 36, 43, 54, Paper I & DA3+dM4e  & close       & S  & 54        & see Appendix \ref{s:t2notes} \\
           &           &                            & DA3+dM4.5 & close       & IP & 8         & \\
           &           &                            & dM2.5--4  & $\lesssim4$ & IP & this work & \\[6pt]

1027-039   & candidate & this work & dM2--7 & $\lesssim4$ & IP & this work & see Appendix \ref{s:t2notes} \\[6pt]

1033+464   & confirmed & 8, 14, 15, 36, 43 & DA2+dM4.5 & $\leq0.5$    & IP & 8         & \nodata  \\
           &           &                   & dM2.5--4  & $\lesssim4$  & IP & this work & \\
           &           &                   & WD+dM4--5 & $\lesssim10$ & IP & 14        & \\[6pt]

1036-204   & tentative & Paper II, this work & \nodata & $\lesssim4$ & IP & this work & see Appendix \ref{s:t2notes} \\[6pt]

1037+512   & confirmed & 36, Paper I & WD+dM      & \nodata     & S  & 36        & \nodata \\
           &           &             & dM2.5--4.5 & $\lesssim4$ & IP & this work & \\[6pt]
   
1042-690   & confirmed & 8, 44 & DA2+dM4.5 & close       & IP & 8         & see Appendix \ref{s:t2notes} \\
           &           &       & dM4--5    & $\lesssim4$ & IP & this work & \\[6pt]

1049+103   & confirmed & 8, 15, 36, 43 & DA2+dM4 & $\leq0.5$   & IP & 8         & \nodata \\
           &           &               & \nodata & $\lesssim4$ & IP & this work & \\[6pt]

1055-072   & tentative & this work & \nodata & $\lesssim4$ & IP & this work & see Appendix \ref{s:t2notes} \\[6pt]

1101+364   & tentative & this work & \nodata & $\lesssim4$ & IP & this work & see Appendix \ref{s:t2notes} \\[6pt]

1104+044   & confirmed & this work & WD+dK0--4 & $\approx3$ & IP & this work & see Appendix \ref{s:t2notes} \\[6pt]

1106+316   & candidate & Paper I & dK5--M0 & $\lesssim4$ & IP & this work & \nodata \\[6pt]

1106-211   & tentative & Paper I & \nodata & $\lesssim4$ & IP & this work & see Appendix \ref{s:t2notes} \\[6pt]

1123+189   & confirmed & 8, 15, 43, 56, Paper I & DA1+dM3 & 1.3         & IP & 8         & \nodata \\
           &           &                        & dM2--3  & $\lesssim4$ & IP & this work & \\
           &           &                        & DA4+dM  & \nodata     & S  & 56        & \\[6pt]

1126+185   & tentative & 15, 49 & DC8+dG--K & \nodata     & S  & 49        & see Appendix \ref{s:t2notes} \\
           &           &        & dK1--4    & $\lesssim4$ & IP & this work & \\[6pt]

1132-298   & confirmed & 43, 53 & DA+dM4  & close       & S  & 53        & see Appendix \ref{s:t2notes} \\
           &           &        & dK7--M2 & $\lesssim4$ & IP & this work & \\[6pt]

1133+358   & confirmed & 17, 43, 49, Paper I & DC+dM4.5e & \nodata     & S  & 49        & \nodata \\
           &           &                     & dM2.5--4  & $\lesssim4$ & IP & this work & \\[6pt]

1136+667   & confirmed & 21, 23, 43, 63, Paper I & DAO+dMe & close       & S  & 23        & see Appendix \ref{s:t2notes} \\
           &           &                         & DAO+dK7 & close       & S  & 63        & \\
           &           &                         & dK6--M0 & $\lesssim4$ & IP & this work & \\[6pt]

1141+504   & confirmed & 15, 60 & DA+dM4e   & close       & S  & 60        & \nodata \\
           &           &        & dK3--M3.5 & $\lesssim4$ & IP & this work & \\[6pt]

1147+371   & candidate & this work & dK5--M2.5 & $\lesssim4$ & IP & this work & \nodata \\[6pt]

1156+129   & tentative & Paper I & dK6--M2.5 & $\lesssim4$ & IP & this work & see Appendix \ref{s:t2notes} \\[6pt]

1201+437   & tentative & 12, 15, 43, Paper I & DC+dMe  & \nodata     & \nodata & 12        & see Appendix \ref{s:t2notes} \\
           &           &                     & \nodata & $\lesssim4$ & IP      & this work & \\[6pt]

1210+464   & confirmed & 8, 15, 36, 43, Paper I & DA2+dM2 & $\leq0.5$   & IP & 8         & \nodata \\
           &           &                        & dM0--2  & $\lesssim4$ & IP & this work & \\[6pt]

1213+528   & confirmed & 34, 43, 50 & DA+dM2    & close       & S  & 34        & see Appendix \ref{s:t2notes} \\
           &           &            & dM3--5    & $\lesssim4$ & IP & this work & \\
           &           &            & DA+dM4.5e & \nodata     & P  & 50        & \\[6pt]

1214+032   & confirmed & 43, 55, 62, Paper I & DA+sdM3 & 2           & IS & 55, 62    & \nodata \\
           &           &                     & dM2--3  & $\lesssim4$ & IP & this work & \\[6pt]

1218+497   & confirmed & 60, Papers I \& III & WD+dM4  & 0.302       & IP & Paper III & \nodata \\
           &           &                     & DA+dM4e & close       & S  & 60        & \\
           &           &                     & dK4--M2 & $\lesssim4$ & IP & this work & \\[6pt]

1224+309   & confirmed & 15, 46, Paper I & DA+dM4+ & close & PS & 46 & see Appendix \ref{s:t2notes} \\[6pt]

1246+299   & candidate & this work & dK3--M0 & $\lesssim4$ & IP & this work & \nodata \\[6pt]

1247-176   & confirmed & 32, 33, Paper I & DA+dMe & close       & S  & 32, 33    & see Appendix \ref{s:t2notes} \\
           &           &                 & dM1--3 & $\lesssim4$ & IP & this work & \\[6pt]

J1255+258  & confirmed & 10, 43 & sdO+G5 III & close          & S  & 10, 25    & see Appendix \ref{s:t2notes} \\
           &           &        & \nodata    & $\lesssim0.05$ & I  & 5         & \\
           &           &        & G3--K0 III & $\lesssim4$    & IP & this work & \\[6pt]

1254-133   & candidate & this work & dM0--5 & $\lesssim4$ & IP & this work & \nodata \\[6pt]

1305+018   & confirmed & 4, this work & dM1--3.5 & $\lesssim2$ & IP & this work & see Appendix \ref{s:t2notes} \\[6pt]

1307-141   & candidate & Paper I & dK7--M2.5 & $\lesssim4$ & IP & this work & \nodata \\[6pt]

1310-305   & tentative & this work & \nodata & $\lesssim4$ & IP & this work & see Appendix \ref{s:t2notes} \\[6pt]

1314+293   & confirmed & 36, 39 & DAwk+dM3.5e & 3           & IPS & 39        & see Appendix \ref{s:t2notes} \\
           &           &        & dM0.5--3    & $\lesssim4$ & IP  & this work & \\[6pt]

1319-288   & confirmed & 33 & DA+dM & \nodata & S & 33 & \nodata \\[6pt]

1330+793   & confirmed & Paper I & dM0--2 & $\approx3.9$ & IP & this work & see Appendix \ref{s:t2notes} \\[6pt]

1333+487   & confirmed & 16, 43, Papers I \& III & WD+dM5 & 2.947                   & IP & Paper III & \nodata \\
           &           &                         & DB+dM  & ``unresolved visually'' & PS & 16        & \\
           &           &                         & dM2--4 & $\lesssim4$             & IP & this work & \\[6pt]

J1340+604  & confirmed & 43, 60, Paper III & WD+dM4  & $\lesssim0.025$ & IP & Paper III & see Appendix \ref{s:t2notes} \\
           &           &                   & DA+dM3e & close           & S  & 60        & \\[6pt]

1401+005   & candidate & this work & dK5+ & $\lesssim4$ & IP & this work & \nodata \\[6pt]

1412-049   & confirmed & Papers I \& III & WD+dM0 & 3.508 & IP & Paper III & \nodata \\[6pt]

1412-109   & confirmed & this work & K8--M5 III & $\lesssim4$ & IP & this work & see Appendix \ref{s:t2notes} \\[6pt]

1415+132   & confirmed & 11, 19 & DA+dM3+ & close       & S  & 11        & \nodata \\
           &           &        & dK3--M2 & $\lesssim4$ & IP & this work & \\[6pt]

1424+503   & confirmed & 43, 59 & DA2+dM    & close       & S  & 59        & see Appendix \ref{s:t2notes} \\
           &           &        & dK7--M2.5 & $\lesssim4$ & IP & this work & \\[6pt]

1433+538   & confirmed & 15, 43, 60, Paper III & DA+dM4    & close           & S  & 60        & see Appendix \ref{s:t2notes} \\
           &           &                       & WD+dM5    & $\lesssim0.025$ & IP & Paper III & \\
           &           &                       & DA2+dM4.5 & $\leq0.5$       & IP & 8         & \\[6pt]

1435+370   & confirmed & Papers I \& III & WD+dM2.5 & 1.251 & IP & Paper III & \nodata \\[6pt]

1436-216   & confirmed & 78, Paper I & DA+dM    & \nodata     & S  & 78        & \nodata \\
           &           &             & dM2--3.5 & $\lesssim4$ & IP & this work & \\[6pt]

1443+336   & confirmed & 15, 36, Papers I \& III & WD+dM2.5 & 0.679       & IP & Paper III & see Appendix \ref{s:t2notes} \\
           &           &                         & dK4--M3  & $\lesssim4$ & IP & this work & \\[6pt]

1458+171   & confirmed & Papers I \& III & WD+dM5   & $\lesssim0.025$ & IP & Paper III & \nodata \\
           &           &                 & dM4--5.5 & $\lesssim4$     & IP & this work & \\[6pt]

1501+300   & tentative & this work & \nodata & $\lesssim4$ & IP & this work & see Appendix \ref{s:t2notes} \\[6pt]

1502+349   & confirmed & Papers I \& III & WD+dM5 & 1.913 & IP & Paper III & \nodata \\[6pt]

1504+546   & confirmed & Papers I \& III & DA+dMe & \nodata         & S  & 65        & \nodata \\
           &           &                 & WD+dM4 & $\lesssim0.025$ & IP & Paper III & \\
           &           &                 & dM1--4 & $\lesssim4$     & IP & this work & \\[6pt]

1517+502   & confirmed & 37, 43, 60, Papers I \& III & DA+dCe & close           & S  & 60        & see Appendix \ref{s:t2notes} \\
           &           &                             & WD+dC  & $\lesssim0.025$ & IP & Paper III & \\
           &           &                             & dM9--L & $\lesssim4$     & IP & this work & \\[6pt]

1522+508   & confirmed & 60, Paper I & DA+dM4e & close       & S  & 60        & \nodata \\
           &           &             & dK5--M3 & $\lesssim4$ & IP & this work & \\[6pt]

1541-381   & tentative & this work & $\leq$dM3 & $\lesssim4$ & IP & this work & see Appendix \ref{s:t2notes} \\[6pt]

1558+616   & confirmed & 43, Papers I \& III & WD+dM4.5  & 0.715       & IP & Paper III & see Appendix \ref{s:t2notes} \\
           &           &                     & dK5--M3.5 & $\lesssim4$ & IP & this work & \\[6pt]

1603+125   & confirmed & Papers I \& III & WD+dK3 & $\lesssim0.025$ & IP & Paper III & see Appendix \ref{s:t2notes} \\
           &           &                 & dK0--4 & $\lesssim4$     & IP & this work & \\[6pt]

1608+118   & confirmed & 8 & DA2+dM3  & 3.0         & IP & 8         & see Appendix \ref{s:t2notes} \\
           &           &   & dM0--2.5 & $\lesssim4$ & IP & this work & \\[6pt]

1610+383   & confirmed & Paper I, this work & WD+dM1--5 & $\approx4$ & IP & this work & see Appendix \ref{s:t2notes} \\[6pt]

1619+525   & confirmed & Papers I \& III & WD+dM+dM & 0.466, 2.596 & IP & Paper III & see Appendix \ref{s:t2notes} \\
           &           &                 & dK6--M0  & $\lesssim4$  & IP & this work & \\[6pt]

1619+414   & confirmed & 43, 74, Papers I \& III & WD+dM5   & 0.231       & IP & Paper III & \nodata \\
           &           &                         & dM2--4.5 & $\lesssim4$ & IP & this work & \\
           &           &                         & DA+dM    & \nodata     & S  & 74        & \\[6pt]

1622+323   & confirmed & 11, 15, 36, 43, Papers I \& III & WD+dM1 & 0.094       & IP & Paper III & see Appendix \ref{s:t2notes} \\
           &           &                                 & dM0--1 & $\lesssim4$ & IP & this work & \\[6pt]

1631+781   & confirmed & 6, 8, 43, 64, Papers I \& III & WD+dM3+dM3  & 0.007, 0.163 & IP & Paper III & see Appendix \ref{s:t2notes} \\
           &           &                               & DA+dM2--5   & close        & IS & 6         & \\
           &           &                               & DA2+dM4--5e & close        & S  & 64        & \\
           &           &                               & DA1+dM3     & $\leq0.5$    & IP & 8         & \\
           &           &                               & dM1--3      & $\lesssim4$  & IP & this work & \\[6pt]

1632-227.1 & candidate & this work & late dK -- late dM & $\lesssim4$ & IP & this work & see Appendix \ref{s:t2notes} \\[6pt]

1634-573   & confirmed & 48, 72 & DO+dK0  & 2.35         & S  & 48, 72    & see Appendix \ref{s:t2notes} \\
           &           &        & dG9--K5 & $\lesssim4$  & IP & this work & \\[6pt]

1643+143   & confirmed & 8, 11, 31, 36, 58, Paper I & DA2+dM2   & $\leq0.5$   & IP & 8         & see Appendix \ref{s:t2notes} \\
           &           &                            & dK5--M0.5 & $\lesssim4$ & IP & this work & \\[6pt]

1646+062   & confirmed & 15, 36, 43, Paper III & WD+dM3.5 & 0.163       & IP & Paper III & \nodata \\
           &           &                       & dK5--M2  & $\lesssim4$ & IP & this work & \\[6pt]

1654+160   & confirmed & 8, Paper I & DB2+dM4.5 & 3.50        & IP & 8         & see Appendix \ref{s:t2notes} \\
           &           &            & dM2--5    & $\lesssim4$ & IP & this work & \\[6pt]

J1711+667  & confirmed & 14, Paper I & WD+dM5--5.5        & $\approx2.5$ & IP & 14        & see Appendix \ref{s:t2notes} \\
           &           &             & late dK -- late dM & $\lesssim4$  & IP & this work & \\[6pt]

1717-345   & confirmed & 28, Paper I & DA+M3.5e           & \nodata      & S  & 28        & see Appendix \ref{s:t2notes} \\
           &           &             & late dK -- late dM & $\lesssim4$  & IP & this work & \\[6pt]

J1820+580  & candidate & 14 & WD+dM5.5--6 & $\lesssim10$ & IP & 14 & see Appendix \ref{s:t2notes} \\[6pt]

1833+644   & confirmed & 43, 73 & DA+dM    & \nodata     & S  & 73        & \nodata \\
           &           &        & dM1--4.5 & $\lesssim4$ & IP & this work & \\[6pt]

1845+019   & candidate & 7, 41 & WD+dM & $\approx3$ & IP & 7 & see Appendix \ref{s:t2notes} \\[6pt]

1844-654   & candidate & this work & dM0--2 & $\lesssim4$ & IP & this work & \nodata \\[6pt]

1950+279   & candidate & this work & dK4--M1 & $\lesssim4$ & IP & this work & \nodata  \\[6pt]

2009+622   & confirmed & 8, 43, Paper III & DA2+dM5  & close           & IP & 8         & see Appendix \ref{s:t2notes} \\
           &           &                  & WD+dM4.5 & $\lesssim0.025$ & IP & Paper III &  \\[6pt]

J2013+400  & confirmed & 43, 67, 71 & DAO+M3    & close       & S  & 67        & see Appendix \ref{s:t2notes} \\
           &           &            & DAO+M4--5 & close       & S  & 71        & \\
           &           &            & dM2.5--4  & $\lesssim4$ & IP & this work & \\[6pt]

J2024+200  & confirmed & 11, 40, 43 & DA+dM    & \nodata     & S  & 11, 40    & see Appendix \ref{s:t2notes} \\
           &           &            & dM0--4.5 & $\lesssim4$ & IP & this work &  \\[6pt]

2101-364   & candidate & this work & dK5--M0.5 & $\lesssim4$ & IP & this work & \nodata \\[6pt]

2108-431   & tentative & this work & dM0--2.5 & $\lesssim4$ & IP & this work & see Appendix \ref{s:t2notes} \\[6pt]

2118-333   & candidate & this work & dM0--2 & $\lesssim4$ & IP & this work & \nodata \\[6pt]

2131+066   & confirmed & 8, 15, 43, 52, 76 & DO1+dM3          & 0.3         & IP & 8         & see Appendix \ref{s:t2notes} \\
           &           &                   & WD+dM1--3        & 0.3         & IP & 52        & \\
           &           &                   & WD+dK5--M0       & \nodata     & S  & 76        & \\
           &           &                   & mid dK -- mid dM & $\lesssim4$ & IP & this work & \\[6pt]

2133+463   & confirmed & 1, 43, Paper I & dM3--4.5 & $\approx3$ & IP & this work & see Appendix \ref{s:t2notes} \\[6pt]

2151-015   & confirmed & 20, 41, 43, 78, Papers I \& III & WD+dM8   & 1.082       & IP & Paper III & \nodata \\
           &           &                                 & DA6+dM8  & $\leq0.5$   & IP & 8         & \\
           &           &                                 & dM7--8.5 & $\lesssim4$ & IP & this work & \\[6pt]

2154+408   & confirmed & 8, 22 & DA2+dM3.5 & close & IP & 8 & see Appendix \ref{s:t2notes} \\[6pt]

2237+819   & confirmed & 13, 43 & WD+dM3--4 & close       & PS & 13        & see Appendix \ref{s:t2notes} \\
           &           &        & dM3--5    & $\lesssim4$ & IP & this work & \\[6pt]

2256+249   & confirmed & 8, 43, 57, Paper I & DA2+dM4   & close & IP & 8  & see Appendix \ref{s:t2notes} \\
           &           &                    & DA2+M3--5 & close & PS & 57 & \\[6pt]

2311-068   & tentative & this work & \nodata & $\lesssim4$ & IP & this work & see Appendix \ref{s:t2notes} \\[6pt]

2317+268   & confirmed & Papers I \& III & WD+dM3.5 & $\lesssim0.025$ & IP & Paper III & \nodata \\
           &           &                 & dM2--5   & $\lesssim4$     & IP & this work & \\[6pt]

2318-137   & tentative & this work & dM1--4 & $\approx3$ & IP & this work & see Appendix \ref{s:t2notes}\\[6pt]

2326-224   & confirmed & Paper I, this work & dM1--4 & $\approx4$ & IP & this work & see Appendix \ref{s:t2notes} \\[6pt]

2336-187   & tentative & Paper I & \nodata & $\approx4$ & IP & this work & see Appendix \ref{s:t2notes}
\enddata
\tablerefs{
(1) \citet{bakos02}, 
(2) \citet{barstow01},
(3) \citet{bruch98},
(4) \citet{cheselka93}, 
(5) \citet{ciardullo99},
(6) \citet{cooke92}, 
(7) \citet{debes05b},
(8) \citet{farihi05a},
(9) \citet{feibelman86},
(10) \citet{feibelman83}, 
(11) \citet{finley97},
(12) \citet{fleming93}, 
(13) \citet{boris04}, 
(14) \citet{green00},
(15) \citet{PG86},
(16) \citet{greenstein75},
(17) \citet{greenstein76},
(18) \citet{greenstein84},
(19) \citet{greenstein86},
(20) \citet{greenstein90},
(21) \citet{heber96}, 
(22) \citet{hilwig02}, 
(23) \citet{holberg01}, 
(24) \citet{hutchings95},
(25) \citet{jasn96},
(26) \citet{jeffries96a},
(27) \citet{jeffries96b},
(28) \citet{kawka04},
(29) \citet{kawka07},
(30) \citet{kellett95},
(31) \citet{kidder91},
(32) \citet{kilkenny97},
(33) \citet{koester01},
(34) \citet{lanning82},
(35) \citet{lanning81},
(36) \citet{liebert05},
(37) \citet{liebert94},
(38) \citet{lisker05},
(39) \citet{margon76},
(40) \citet{mason95}, 
(41) \citet{maxted99},
(42) \citet{maxted07},
(43) \citet{MS99},
(44) \citet{morales05},
(45) \citet{mueller87},
(46) \citet{orosz99}, 
(47) \citet{OHL88},
(48) \citet{parsons76},
(49) \citet{putney97}, 
(50) \citet{probst83},
(51) \citet{raymond03},
(52) \citet{reed00}, 
(53) \citet{ruiz90}, 
(54) \citet{saffer93},
(55) \citet{salim03},
(56) \citet{schmidt95},
(57) \citet{SSH95}. 
(58) \citet{schultz96},
(59) \citet{schwartz95}, 
(60) \citet{silvestri06},
(61) \citet{silvestri02},
(62) \citet{silvestri01},
(63) \citet{sing04}, 
(64) \citet{sion95},
(65) \citet{stepanian01}, 
(66) \citet{thor78},
(67) \citet{thor94}, 
(68) \citet{vandenbos26},
(69) \citet{vennes95},
(70) \citet{vennes94},
(71) \citet{vennes99}, 
(72) \citet{wegner79},
(73) \citet{wegner88},
(74) \citet{wegner90b}, 
(75) \citet{wegner87}, 
(76) \citet{wesemael85}, 
(77) \citet{wick77},
(78) \citet{zuckerman03},
(79) \citet{zwitter95}
}
\end{deluxetable}

\clearpage

\begin{deluxetable}{lcccc}
\tablewidth{0pt}
\tablecaption{White Dwarf Statistics 
\label{t:stats}}
\tablehead{
\colhead{        } & \multicolumn{4}{l}{Status in this work:} \\
\colhead{Category} & \colhead{Total} & \colhead{Confirmed} & \colhead{Candidate} & \colhead{Tentative}
}
\startdata
This work totals &                            154 &       105 &    28 &    21 \\
New to this work &                         \phn83 &    \phn46 &    20 &    17 \\
Candidate in Paper I\tablenotemark{a} &    \phn68 &    \phn58 & \phn7 & \phn3 \\
Tentative in Paper I\tablenotemark{b} & \phn\phn3 & \phn\phn1 & \phn1 & \phn1 
\enddata
\tablenotetext{a}{Table 1 in Paper I.}
\tablenotetext{b}{Table 2 in Paper I.}
\end{deluxetable}

\begin{deluxetable}{lrrrrrr}
\tablewidth{0pt}
\tablecaption{Absolute Magnitudes of Low Mass Stars 
\label{t:dists}}
\tablehead{
\colhead{Spectral} & 
\multicolumn{3}{c}{Absolute Magnitude\tablenotemark{a}} &
\multicolumn{3}{c}{Maximum Distance Modulus\tablenotemark{b}} \\
\colhead{Type} & 
\colhead{$M_{J}$} &
\colhead{$M_{H}$} &
\colhead{$M_{K_{\rm s}}$} &
\colhead{$(m-M)_{J}$} &
\colhead{$(m-M)_{H}$} &
\colhead{$(m-M)_{K_{\rm s}}$}
}
\startdata
M0 &  6.45 &  5.82 &  5.62 & 9.35 & 9.28 & 8.67 \\
M1 &  6.72 &  6.11 &  5.88 & 9.08 & 8.99 & 8.41 \\
M2 &  6.98 &  6.38 &  6.14 & 8.82 & 8.72 & 8.16 \\
M3 &  7.24 &  6.68 &  6.40 & 8.56 & 8.42 & 7.90 \\
M4 &  8.34 &  7.80 &  7.50 & 7.46 & 7.30 & 6.80 \\
M5 &  9.44 &  8.88 &  8.54 & 6.36 & 6.22 & 5.76 \\
M6 & 10.18 &  9.54 &  9.15 & 5.62 & 5.56 & 5.15 \\
M7 & 10.92 & 10.29 &  9.89 & 4.88 & 4.81 & 4.41 \\
M8 & 11.14 & 10.46 & 10.05 & 4.66 & 4.64 & 4.24 \\
M9 & 11.43 & 10.71 & 10.26 & 4.37 & 4.39 & 4.04 \\
L0 & 11.72 & 10.90 & 10.32 & 4.08 & 4.20 & 3.98 \\
L1 & 12.00 & 11.20 & 10.57 & 3.80 & 3.90 & 3.73 \\
L2 & 12.29 & 11.28 & 10.58 & 3.51 & 3.82 & 3.72 \\
L3 & 12.58 & 11.54 & 10.86 & 3.22 & 3.56 & 3.44 \\
L4 & 12.87 & 11.68 & 11.00 & 2.93 & 3.42 & 3.30 \\
L5 & 13.16 & 11.99 & 11.29 & 2.64 & 3.11 & 3.01 \\
L6 & 14.31 & 13.13 & 12.25 & 1.49 & 1.97 & 2.05 \\
L7 & 14.45 & 13.27 & 12.51 & 1.35 & 1.83 & 1.79 \\
L8 & 14.58 & 13.34 & 12.62 & 1.22 & 1.76 & 1.68 \\
WD\tablenotemark{c} & 13.19(57) & 13.12(50) & 13.06(48) & 2.61 & 1.98 & 1.24 \\
\enddata
\tablenotetext{a}{From \citet{bessell88,gizis00,kirkpatrick00,hawley02}.}
\tablenotetext{b}{For 2MASS Good detection limits of $J_{\rm lim}=15.8$, $H_{\rm lim}=15.1$, and $K_{\rm s,lim}=14.3$.}
\tablenotetext{c}{Average photometry from the single WD sample used in the Paper I simulation.}
\end{deluxetable}



\end{document}